\documentclass[fleqn, usenatbib]{mnras}

\usepackage[T1]{fontenc}
\usepackage{ae,aecompl}

\usepackage{graphicx}	
\usepackage{amsmath}	
\usepackage{amssymb}	
\usepackage{amsfonts}
\usepackage[flushleft]{threeparttable}
\usepackage{booktabs}
\usepackage{longtable}
\usepackage{pdflscape}
\usepackage{caption}
\usepackage{morefloats}

\newcommand{\herschel}{\emph{Herschel}}
\newcommand{\swift}{\textit{Swift}}
\newcommand{\msun}{M$_{\sun}$}
\newcommand{\mstar}{$M_{\mathrm{star}}$}
\newcommand{\lsun}{L$_{\sun}$}
\newcommand{\mdust}{$M_{\rm dust}$}
\newcommand{\tdust}{$T_{\rm dust}$}

\title[\emph{Herschel}-BAT Sample: AGN-SF Connection]{\emph{Herschel} far-infrared photometry of the Swift Burst Alert
Telescope active galactic nuclei sample of the local universe--III. Global star-forming properties and the lack of a connection to nuclear activity\thanks{{\it Herschel} is an ESA space observatory with science instruments provided by European-led Principal Investigator consortia and with important participation from NASA.}}

\author[T. T. Shimizu]{T. Taro Shimizu$^{1,2}$\thanks{Email: shimizu@mpe.mpg.de}, Richard F. Mushotzky$^1$, Marcio Mel\'endez$^{1,3,4}$, Michael J. Koss$^{5}$, \newauthor Amy J. Barger$^{6,7,8}$, and Lennox L. Cowie$^{8}$\\
$^{1}$Department of Astronomy, University of Maryland, College Park, MD 20742, USA\\
$^{2}$Max-Planck-Institut f\"{u}r extraterrestrische Physik, Postfach 1312, 85741, Garching, Germany\\
$^{3}$NASA Goddard Space Flight Center,  Greenbelt, MD  20771, USA\\
$^{4}$KBRwyle Science, Technology and Engineering Group, 1290 Hercules Avenue  Houston, TX  77058, USA\\ 
$^{5}$Institute for Astronomy, Department of Physics, ETH Zurich, Wolfgang-Pauli-Strasse 27, CH-8093 Zurich, Switzerland\\
$^{6}$Department of Astronomy, University of Wisconsin-Madison, 475 N. Charter Street, Madison, WI 53706, USA\\
$^{7}$Department of Physics and Astronomy, University of Hawaii, 2505 Correa Road, Honolulu, HI 96822, USA\\
$^{8}$Institute for Astronomy, University of Hawaii, 2680 Woodlawn Drive, Honolulu, HI 96822, USA}

\date{Accepted XXX. Received YYY; in original form ZZZ}

\pubyear{2016}

\begin{document}
\label{firstpage}
\pagerange{\pageref{firstpage}--\pageref{lastpage}}
\maketitle

\begin{abstract}
We combine the \herschel{} \textit{Space Observatory} PACS and SPIRE photometry with archival \textit{WISE} photometry to construct the spectral energy distributions (SED) for over 300 local ($z < 0.05$), ultra-hard X-ray (14--195 keV) selected active galactic nuclei (AGN) from the \swift{} \textit{Burst Alert Telescope} (BAT) 58 month catalogue. Using a simple analytical model that combines an exponentially cut-off powerlaw with a single temperature modified blackbody, we decompose the SEDs into a host-galaxy and AGN component. We calculate dust masses, dust temperatures, and star-formation rates (SFR) for our entire sample and compare them to a stellar mass-matched sample of local non-AGN galaxies. We find AGN host galaxies have systematically higher dust masses, dust temperatures, and SFRs due to the higher prevalence of late-type galaxies to host an AGN, in agreement with previous studies of the \swift/BAT AGN. We provide a scaling to convert X-ray luminosities into 8--1000 \micron{} AGN luminosities, as well as determine the best mid-to-far IR colors for identifying AGN dominated galaxies in the IR regime. We find that for nearly 30 per cent of our sample, the 70 \micron{} emission contains a significant contribution from the AGN ($> 0.5$), especially at higher luminosities ($L_{14-195\,\rm{keV}} > 10^{42.5}$ ergs s$^{-1}$). Finally, we measure the local SFR-AGN luminosity relationship, finding a slope of 0.18, large scatter (0.37 dex), and no evidence for an upturn at high AGN luminosity. We conclude with a discussion on the implications of our results within the context of galaxy evolution with and without AGN feedback.
\end{abstract}




\section{Introduction}
Ever since the discoveries that the large scale properties of galaxies are related to the mass of the supermassive black holes (SMBH) they host \citep{Magorrian:1998qv, Ferrarese:2000gf, Gebhardt:2000xy, Marconi:2003ve, Haring:2004ly, Gultekin:2009ul, McConnell:2013fk, Kormendy:2013fj}, studies have intensely focused on finding the link between the evolution and growth of both the galaxy and SMBH. Theoretical arguments \citep[eg.][]{Silk:1998qf}, as well as cosmological simulations  \citep[e.g][]{Croton:2006kx, Bower:2006gf} support the idea that SMBHs, while accreting material in their active galactic nuclei (AGN) phase, can affect both their own growth and the growth of their host galaxy through a mechanism that shuts down accretion and star formation (i.e. negative feedback). 

Early evidence backing this theory seemed to be found in the observation that the SFR density and accretion rate density of the universe evolve similarly with redshift  \citep{Boyle:1998vn, Franceschini:1999ys, Silverman:2008zl, Silverman:2009lq, Aird:2010fr, Merloni:2013zr}. Further, AGN host galaxies were found to predominantly lie in the ``green valley''  of the color-magnitude diagram, in between the blue, star-forming galaxies and ``red-and-dead'' quiescent ones \citep{Martin:2007yg, Nandra:2007rz, Silverman:2008zl, Schawinski:2010zr, Koss:2011vn}, suggesting that nuclear activity plays a role in transitioning its host galaxy from star-forming to quiescence. 

While suggestive, optical colors can be strongly affected by extinction due to dust that might move a galaxy from ``blue'' to ''green'' \citep{Cowie:2008qy, Cardamone:2010uq}. Other measures of the SFR including emission line, ultra-violet (UV), and mid-infrared (MIR) luminosities that are extensively used for normal, non-AGN galaxies suffer from varying degrees of contamination in the presence of an AGN, especially unabsorbed ($N_{\rm H} < 10^{22}$ cm$^{-2}$) Type I AGN.

This problem was solved with the advent of far-infrared (FIR; $\lambda = 40-500$ \micron) telescopes, including the \textit{Infrared Astronomical Satellite} (\textit{IRAS}), the \textit{Infrared Space Observatory} (\textit{ISO}), the \textit{Spitzer Space Telescope}, and most recently the \herschel{} \textit{Space Observatory} \citep{Pilbratt:2010rz}. FIR emission remains the ideal waveband to study star formation in AGN host galaxies for two reasons: (1) the rapidly declining spectral energy distribution (SED) associated with AGN heated dust in the FIR regime \citep{Fritz:2006yq, Netzer:2007ve, Mullaney:2011yq, Shi:2014qr} ensures little AGN contamination, and (2) thermal emission of large grains heated by the ionizing emission of recently formed massive stars \citep{Lonsdale-Persson:1987uq, Devereux:1990yq} creates a strong FIR ``bump'' and provides a reliable measure of the recent (10--100 Myr) SFR \citep{Kennicutt:1998kx}.

Even though the FIR can provide a reliable SFR tracer, individual investigations into the connection between AGN activity and star-formation have produced conflicting and even contradictory results. Some studies \citep{Page:2012mz, Barger:2015ly} have reported lowered SFRs for the highest luminosity AGN, indicative of a suppression of star formation due to AGN feedback. Others have found weak or flat relationships \citep{Diamond-Stanic:2012rw, Mullaney:2012gf, Rovilos:2012wd, Rosario:2012fr, Azadi:2015ve, Stanley:2015qy}, especially at more moderate AGN luminosities. Finally, some find an overall positive correlation \citep{Lutz:2008ph, Netzer:2009lr, Rovilos:2012wd, Rosario:2012fr, Chen:2013uq, Dai:2015gf}.

Many of the studies may however suffer from one or more problems that include small number statistics either due to a small area survey or low sensitivity \cite[e.g.][]{Harrion:2012lj}, an inaccurate scaling from monochromatic luminosities to a total SFR, and source confusion due to the relatively large beam size of \herschel, especially at longer wavelengths. Further, as \citet{Barger:2015ly} pointed out, stacking and averaging analyses can overestimate the level of star formation taking place in the bulk of the X-ray sample because the true SFR distribution is unknown. 

In this paper, we examine the AGN-star formation relationship in the low redshift universe. We have performed a \herschel{} survey of a sample of AGN from the \textit{Swift} Burst Alert Telescope 58 month catalogue, which selected sources over the entire sky in the 14--195 keV energy range. At this high energy, we are nearly unbiased with respect to any host galaxy properties including SFR, as well as obscuration below a hydrogen column density ($N_{\rm H}$) of 10$^{24}$ cm$^{-2}$. The \swift/BAT AGN sample has been extensively studied ever since the release of the first catalogues. Most relevant to our work, \citet{Koss:2011vn} performed an optical survey of over 100 \swift/BAT AGN using combined \textit{Sloan Digital Sky Survey} (SDSS) and \textit{Kitt Peak National Observatory} imaging to study their host galaxy properties. They found that the \swift/BAT AGN are dominated by massive (\mstar $> 10^{9.5}$ \msun) spirals with bluer colors than a mass-matched sample of non-AGN galaxies. \citet{Koss:2011vn} further found that AGN showed enhanced FIR luminosities, using 90 \micron{} photometry from \textit{Akari}. 

With 5 band imaging with \herschel{} along with archival photometry from the \textit{Wide-field Infrared Survey Explorer} (\textit{WISE}), we have constructed MIR-FIR SEDs for over 300 AGN, allowing us to accurately measure the SFR in a robust manner. Because of the low-redshift nature of our sources, every AGN is easily able to be identified and there is no concern over source confusion. These SEDs not only allow us to calculate SFRs but also to test varying SED decomposition methods and models, as well as measure other properties of the host galaxies, such as the dust mass and dust temperature, to compare with non-AGN samples. With a high detection rate (95, 83, 86, 72, and 46 per cent at 70, 160, 250, 350, and 500 \micron), our measured properties from the SEDs are well constrained, removing much of the uncertainty due to censoring. 

The paper is organized as follows: in Section 2 we briefly describe the \herschel-BAT AGN sample; in Section 3 we detail our \textit{WISE} and \herschel{} datasets. Section 4 introduces our non-AGN comparison sample. Section 5 qualitatively examines the SEDs and determines the average SED as a function of AGN luminosity, while Section 6 outlines our fitting methods. Sections 7 and 8 then discuss our results, compares with other studies, and concludes the paper. Throughout, we use a cosmology with $H_{0}=70$ km s$^{-1}$ Mpc$^{-1}$, $\Omega_{\rm M}=0.3$, and $\Omega_{\Lambda} = 0.7$ to calculate luminosity distances from redshifts. 

\section{\herschel-BAT Sample}
Because \swift/BAT continuously monitors the entire sky in the energy range 14--195 keV for gamma ray bursts, it simultaneously provides an all-sky survey at ultra high X-ray energies. This allows for the creation of complete catalogues with increasing sensitivity the longer \swift/BAT remains in operation. Given the extreme environments necessary to produce strong 14--195 keV emission, the majority of sources in the \swift/BAT catalogues are AGN.

We chose 313 AGN from the parent sample of $\sim720$ AGN detected in the 58 month catalogue\footnote{\url{https://swift.gsfc.nasa.gov/results/bs58mon}} after imposing a redshift cutoff of $z<0.05$  and excluding Blazars and BL Lac objects to form our \herschel-BAT AGN sample. With a mean redshift of $\left<z\right> = 0.025$, our AGN sample provides a comprehensive view of the properties of AGN host galaxies in the local universe. Our selection at ultra high X-ray energies further removes biases and selection effects due to host galaxy contamination and obscuration \citep{Mushotzky:2004gf} that can influence samples at other wavelengths.   

The demographics of our sample are nearly evenly split between Type I (43 per cent) and Type II (53 per cent) with the remaining 4 per cent (5 objects) either a Low-Ionization Nuclear Emission-line Region (LINER) or unclassified. The \herschel-BAT sample spans nearly four orders of magnitude in 14--195 keV luminosity ($10^{41} < L_{\rm 14-195\,keV} < 10^{45}$ ergs s$^{-1}$) allowing for a robust determination of the connection between AGN strength and SFR. For a complete listing of our sample with names, luminosity distances, redshifts, and AGN type, we point the reader to \citet{Melendez:2014yu} or \citet{Shimizu:2016qy}.

\section{Data}
\subsection{\herschel{} Photometry}
293 of the \herschel{}-BAT AGN were observed with \herschel{} as part of a Cycle 1 open time program (OT1\_rmushotz\_1, PI: Richard Mushotzky). The remaining 20 sources were part of other programs with public archival data. \herschel{} observed all 313 AGN using both the Photoconductor Array Camera and Spectrometer \citep[PACS;][]{Poglitsch:2010fp} and Spectral and Photometric Imaging Receiver \citep[SPIRE;][]{Griffin:2010sf} producing images in five wavebands: 70, 160, 250, 350, and 500 \micron.

Detailed descriptions of the data reduction process and photometric flux extraction can be found in \citet{Melendez:2014yu} for PACS and \citet{Shimizu:2016qy} for SPIRE. The following is a short description of the flux extraction procedure. We measured fluxes at each waveband directly from the images using aperture photometry with a concentric annulus to define the local background. We applied aperture corrections for sources where we used a point source aperture as defined in the respective PACS and SPIRE data reduction guides. Fluxes for sources that were unresolved at all three wavebands in SPIRE were determined using the SPIRE Timeline Fitter within the \herschel{} Interactive Processing Environment. All fluxes have a signal-to-noise ratio of at least 5, otherwise we provided a 5$\sigma$ upper limit. 

In \citet{Shimizu:2016qy}, using the SPIRE flux ratios, we found 6 radio-loud objects, which have a large contribution to their FIR emission from synchrotron radiation due to their jets. These sources are Pictor A, 3C 111.0, 3C 120, 2MASX J23272195+1524375, PKS2331-240, and [HB89] 0241+622. For this work we remove these sources from all analysis; however, we still provide figures of their SED fits in Appendix D and list their best-fit parameters in Table~\ref{tab:c12_params}. 

Also in \citet{Shimizu:2016qy}, we flagged objects with a ``D'' that likely had strong contamination due to a nearby source. For these sources, we treat the SPIRE fluxes as upper limits. For a small number of very nearby galaxies with large angular size (NGC 2655, NGC 3718, NGC 4939, NGC 4941, NGC 5033, and NGC 6300), we realized that the PACS FOV of our observations did not encompass the entirety of the galaxy leading to possible underestimations of the 70 and 160 \micron{} flux densities. Therefore, for this work, we chose to conservatively increase the uncertainty to 50 per cent.

\subsection{\textit{WISE} Photometry}
To extend the SEDs into the MIR, we supplemented our \herschel{} data with archival \textit{Wide-field Infrared Survey Explorer} \citep[WISE;][]{Wright:2010fk} photometry. WISE performed a broadband all-sky survey at 3.4 (W1), 4.6 (W2), 12 (W3), and 22 (W4) \micron{} with angular resolution comparable to \herschel/PACS at 70 \micron{} for W1, W2, and W3 and 160 \micron{} for W4. We queried the AllWISE catalogue through the NASA/IPAC Infrared Science Archive\footnote{\url{http://irsa.ipac.caltech.edu/frontpage/}} to search for coincident sources within 6\arcsec. Counterparts were found for all but one AGN (Mrk 3) at every waveband. The catalogue only contained W1 and W2 fluxes for Mrk 3 due to differences in the depth of coverage for W1/W2 and W3/W4, therefore Mrk 3 is not included as part of the sample in this work.

The AllWISE catalogue provides magnitudes determined using multiple extraction methods. We consider only the profile-fitting magnitudes (\textit{w}N\textit{mpro} where N is 1, 2, 3, or 4) and the elliptical aperture magnitudes (\textit{w}N\textit{gmag}). Profile-fitting magnitudes were determined by fitting the position dependent point spread function using deblending procedures when necessary to decompose overlapping sources. The \textit{w}N\textit{mpro} magnitudes therefore are only relevant for unresolved sources.

If a WISE source is associated with a source in the \textit{Two Micron All Sky Survey} (2MASS) Extended Source Catalog (XSC), then \textit{w}N\textit{gmag} magnitudes were also measured using an elliptical aperture with the same shape from the XSC and sizes scaled given the larger WISE beam. Thus, \textit{w}N\textit{gmag} magnitudes are more appropriate for extended sources. For details of all of the WISE magnitude measurements we point the reader to the All-Sky Release Explanatory Supplement\footnote{\url{http://wise2.ipac.caltech.edu/docs/release/allsky/expsup/}}.

Given the low-redshift nature of our sample, 294/313 galaxies are in the 2MASS XSC and using only the \textit{w}N\textit{mpro} magnitudes would severely underestimate the flux for many of them. To decide which magnitude to include in the SED for each source, we used the reduced $\chi^{2}$ value (\textit{w}N\textit{rchi2}) from the profile-fitting. If \textit{w}N\textit{rchi2}$<3$ then we chose the \textit{w}N\textit{mpro} magnitude, otherwise \textit{w}N\textit{gmag} was chosen. For 55/313 sources, we used the W1 profile-fitting magnitudes; for 129/313 sources, we used the W2 profile-fitting magnitudes; for 210/313 sources, we used the W3 profile-fitting magnitudes; and for 267/313 sources, we used the W4 profile-fitting magnitudes. The increasing numbers of sources for which we use the profile-fitting magnitudes is due to the increasing FWHM of the WISE point spread function which changes from 6\arcsec at W1 to 12\arcsec at W4.

\section{Comparison Sample}
To test whether the AGN has any effect on the star-forming properties of their host galaxies, we need samples of galaxies which contain little evidence for nuclear activity. These samples also need to occupy the same redshift range to mitigate against evolutionary effects and have been observed with nearly the same instruments so the properties can be compared on an equal level. Three low-redshift samples exist that satisfy these constraints. They are the  \herschel{} \textit{Reference Survey} \cite[HRS;][]{Boselli:2010fj}, \textit{Key Insights on Nearby Galaxies: a Far-Infrared Survey with Herschel} \cite[KINGFISH;][]{Kennicutt:2011vn}, and the \herschel{} \textit{Stripe 82 Survey} \citep[HerS;][]{Viero:2014jk}. While the KINGFISH sample has been observed at all of the same wavelengths as the \herschel-BAT AGN, the sample was selected in a heterogeneous manner and only contains 61 galaxies, which does not allow for stellar mass matching. The HerS sample is numerous and covers a large range in mass, but was only observed with SPIRE. Further, the catalog was built based on SPIRE detections, which imposes a selection effect associated with the SFR. We therefore chose the HRS as our comparison sample to the \herschel-BAT AGN.

The HRS was a guaranteed time key \herschel{} program dedicated to studying the dust content of ``normal'' galaxies. The 323 galaxy sample is volume limited ($15< D < 25$ Mpc) to avoid distance effects and K-band flux limited to avoid selection effects due to dust and provide a representative population of local galaxies. The size of the HRS, as well as the local nature make it an ideal sample to compare to the \herschel-BAT galaxies.

The HRS galaxies also were imaged using both PACS and SPIRE, although the 100 \micron{} filter was used instead of the 70 \micron{} filter. \citet{Cortese:2014qq} and \citet{Ciesla:2012lq} measured the PACS and SPIRE flux densities respectively using similar techniques as the \herschel-BAT galaxies. WISE 12 and 22 \micron{} photometry for HRS were provided in \citet{Ciesla:2014qy}. The available data for the HRS SEDs are nearly identical to our sample. 

\begin{figure}
\includegraphics[width=\columnwidth]{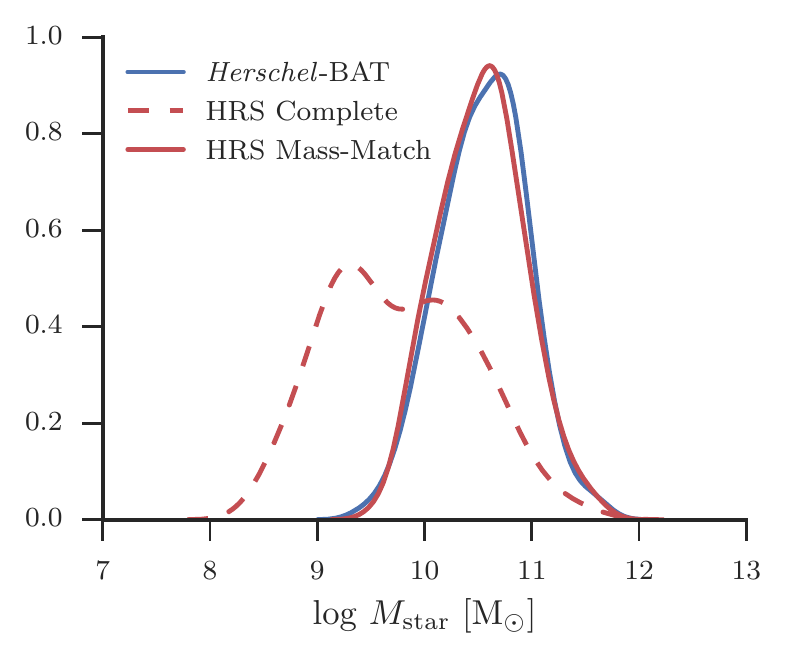}
\caption{Kernel Density Estimates (KDE) of the \mstar{} distribution for the BAT AGN (blue), full HRS sample (red), and mass-matched HRS sample (red dashed). The \herschel-BAT AGN probe a higher \mstar{} galaxy population than the full samples; however, with mass-matching we can reproduce the stellar-mass distribution of the \herschel-BAT AGN with non-AGN. A color version is available in the online publication. \label{fig:mstar_dist}}
\end{figure}

The only issue in comparing the HRS galaxies to the \herschel-BAT AGN concerns the stellar mass (\mstar) distribution. Because the near-infrared is most strongly effected by the mass of the older stellar population, the HRS K-band selection produces a \mstar{} distribution that is representative of the naturally occurring \mstar{} distribution. However, many recent studies have shown that AGN prefer high \mstar{} galaxies \citep[e.g.][]{Schawinski:2010zr,Xue:2010fj}, a feature that is also found in the BAT AGN \citep{Koss:2011vn}. Figure~\ref{fig:mstar_dist} displays the \mstar{} distributions for the \herschel-BAT and HRS sample (dashed line) represented using a Kernel Density Estimate (KDE)\footnote{Instead of the standard histogram, we choose to visualize distributions in this work with KDEs. KDEs are constructed by first representing each data point in the sample by a specific kernel which are then summed together. We use a Gaussian kernel with a width according to ``Scott's Rule''  \citep{Scott:1992xy}, $width=N^{-1/5}$, where $N$ is the number of data points. The appearance of a histogram can change based on the number of bins, bin width, and bin centers whereas a KDE is only dependent on the kernel width.} The \herschel-BAT galaxies have an average \mstar{} of 10.6 \msun, whereas the average \mstar{} for HRS is 9.8 \msun, nearly a ten-fold difference. 

To account for possible stellar mass effects, we decided to produce a mass-matched HRS sample in the following way. Stellar masses for the \herschel-BAT AGN were calculated based on the AGN-subtracted \textit{ugriz} photometry from \citet{Koss:2011vn} and the $g-i$ calibrated stellar mass relation from \citet{Zibetti:2009jf}. This is the same method that stellar masses were calculated for the HRS sample, given in \citet{Cortese:2012fj}. For each of the 122 \herschel-BAT AGN which have stellar mass estimates, we chose all HRS galaxies which have a stellar mass within 0.15 dex of the \herschel-BAT mass. We then randomly selected an HRS galaxy from this pool. Because high-mass galaxies are relatively rare in the HRS sample, some matches are duplicated. We find that 97/122 of the matched HRS galaxies are unique, representing 80 per cent of the sample, so the duplicating effects should not be large. The red solid line in Figure~\ref{fig:mstar_dist} now shows the mass-matched HRS sample, which overlaps with the \herschel-BAT mass distribution. Throughout the rest of this paper, we refer to the HRS mass-matched sample as simply the HRS sample unless otherwise noted.
 
\section{The IR SEDs of $z=0$ AGN}\label{sec:ir_seds}

Before fitting the SEDs and comparing to a non-AGN sample, we begin with a qualitative look at the SEDs of the 313 AGN. Figure~\ref{fig:seds} shows the 12--500 \micron{} SED for every source in the \herschel-BAT sample. Immediately noticeable is the varied SED shapes. Some of the SEDs (e.g. 2MASX J07595347+2323241, Cen A, ESO 005-G004, NGC 4051, NGC 4138, NGC 6814) feature a very prominent FIR bump that is recognizable in nearly all star-forming galaxies and is indicative of thermally heated dust from recently formed massive stars, as well as cirrus emission heated by an older population. The general shape is usually well fit by a blackbody modified by a frequency dependent optical depth \citep[e.g.][]{Calzetti:2000fk, Smith:2012fj, Bianchi:2013jk, Symeonidis:2013fe, Cortese:2014qq}.

These star formation dominated objects represent one extreme end of the shapes we observe. The other extreme are SEDs whose shapes seem to peak short-ward of 70 \micron{} and display rapidly falling emission with increasing wavelength. Examples include 2MASX J06561197-4919499, 2MASX J210990996-0940147, ESO 103-035, IC 4329A, MCG -05-23-016, and Mrk 335. Emission in the MIR dominates these SEDs and is likely associated with hotter dust heated by the AGN. MIR colors are commonly used to select AGN samples because they display strong red (i.e. increasing SED with increasing wavelength shortward of 70 \micron) colors compared to non-AGN galaxies \citep{Lacy:2004uq, Donley:2012qy, Stern:2012mz} due to the dust in the obscuring torus heated by the optical and UV emission from the accretion disk. Monochromatic MIR luminosities, especially for sources dominated by an unresolved central component, show near linear correlations with the X-ray luminosity \citep[e.g][]{Lutz:2004fj, Gandhi:2009kx, Asmus:2015qy} providing more evidence that much of the MIR in AGN host galaxies is associated with the AGN. 

The remaining \herschel-BAT AGN display SEDs somewhere in between MIR-dominated and FIR-dominated as a result of competing contributions between the AGN and star formation. The varied shapes of the SED emphasizes the need for SED decomposition, especially at shorter wavelengths, to accurately determine the IR luminosity associated with either star formation or the AGN. 

If star formation suppression occurs at high AGN luminosity then we expect the IR SEDs for the most X-ray luminous sources to resemble the MIR-dominated SEDs, as \citet{Barger:2015ly} observed. To test this, we binned the SEDs by logarithmic 14--195 keV luminosity. Five bins ( $\log\,L_{\rm 14-195\,keV} > 42.5$, $\log\,L_{\rm 14-195\,keV} = 42.5-43.0$, $43.0-43.5$, $43.5-44.0$, and $\log\,L_{\rm 14-195\,keV} > 44.0$) were chosen with 0.5 dex widths to ensure enough sources occupied each one. The number of sources in each bin is 22, 39, 94, 116, and 35. 

We then calculated the median luminosity density for each waveband in the SED within each X-ray luminosity bin . We used the ASURV \citep{Feigelson:1985lr} package, which applies the principles of survival analysis to astronomical data. Information contained in the upper limits can be then included in the measurement of statistical properties of samples without biasing results towards brighter sources that are more likely to be detected. Specifically, ASURV calculates the non-parametric Kaplan-Meier product-limit (KMPL) estimator for a sample distribution. The KMPL estimator is an estimate of the survival function, which is simply 1$-$CDF (cumulative distribution function). Using the KMPL, we calculate for each waveband the median luminosity density (50th percentile) and estimate the uncertainty using the 16th and 84th percentiles. 

\begin{figure*}
\centering
\includegraphics[width=0.75\textwidth]{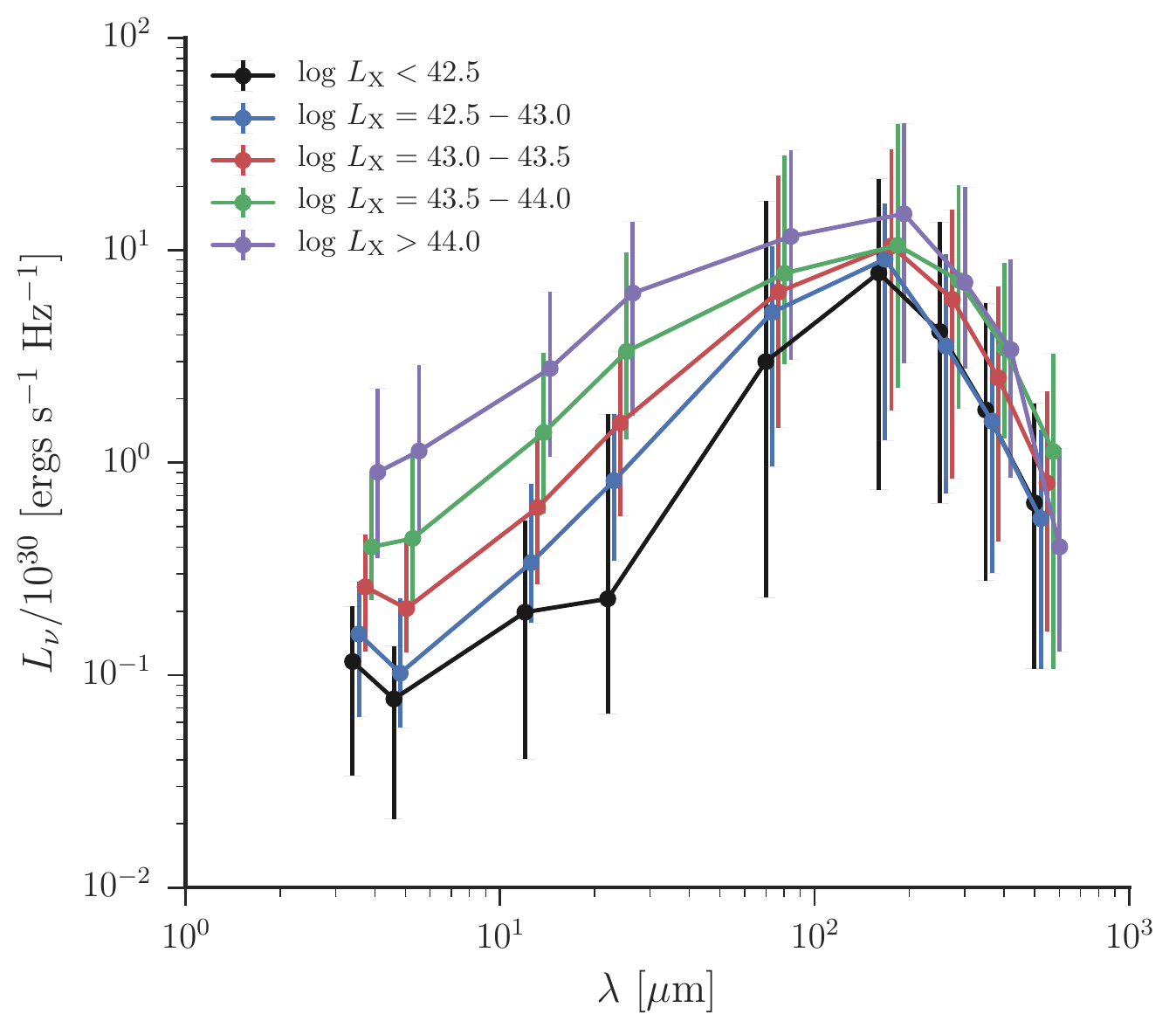}
\caption{Median IR SEDs in units of $10^{30}$ ergs s$^{-1}$ Hz$^{-1}$ for the \herschel-BAT AGN binned by $L_{\rm 14-195\,keV}$. Each point in the SEDs is the median $L_{\nu}$ for the sources in the bin including upper limits calculated using the Kaplan-Meier product-limit estimator. The error bars represent the 68 per cent confidence interval determined from the 16th and 84th percentiles. SEDs have been slightly offset by 0.02 dex in wavelength so the error bars can be clearly shown. \label{fig:avg_sed}}
\end{figure*}

Figure~\ref{fig:avg_sed} shows the median IR SED for sources in each X-ray luminosity bin. From visual inspection, the median SED as a function of X-ray luminosity seems to be driven more by an increasing MIR hot dust component. There does not seem to be any indication that star formation is being suppressed at the highest luminosities. The wavebands that show the largest change with X-ray luminosity are the MIR bands. At wavelengths longer than 160 \micron, the SED shows no significant change, consistent with the correlations we measured in \citet{Melendez:2014yu} and \citet{Shimizu:2016qy}. If we make the assumption that the longer wavelengths are mainly associated with star-formation, then the lack of change indicates star-formation, on timescales of $\sim$100 Myr, is unrelated to X-ray luminosity within the range probed by our sample. However, to definitively test this we need to decompose the SED into star-forming and AGN components to accurately calculate SFRs.

\section{SED Fitting}\label{sec:sed_fit}

\begin{table*}
\centering
\begin{threeparttable}
\captionsetup{font=small,labelfont=bf,labelsep=period}
\caption{Best Fit C12 Model Parameters, Luminosities, and AGN Fractions \label{tab:c12_params}}
\begin{tabular}{lcccccccc}
\toprule 
Name & $\log\,M_{\rm dust}$ &$T_{\rm dust}$ & $\alpha$ & $\lambda_{\rm c}$ & $\log\,L_{\rm IR}$ & $\log\,L_{\rm SF}$ & $\log\,L_{\rm AGN,IR}$ &$f_{\rm AGN}$ \\
 & [\msun] & [K] &  & [\micron] & [\lsun] & [\lsun] & [\lsun] & \\
\midrule
1RXSJ044154.5-082639 & $6.80_{-0.14}^{+0.21}$ & $27.98_{-4.21}^{+2.45}$ & $1.70_{-0.34}^{+0.46}$ &$47.71_{-13.96}^{+18.85}$ & $10.37_{-0.04}^{+0.03}$ & $9.99_{-0.23}^{+0.11}$ & $9.99_{-0.23}^{+0.11}$ & $0.58_{-0.14}^{+0.18}$ \\
1RXSJ045205.0+493248 & $7.47_{-0.10}^{+0.12}$ & $23.27_{-1.61}^{+0.99}$ & $1.44_{-0.46}^{+0.59}$ &$39.95_{-13.09}^{+14.82}$ & $10.48_{-0.04}^{+0.03}$ & $10.19_{-0.08}^{+0.04}$ & $10.19_{-0.08}^{+0.04}$ & $0.50_{-0.10}^{+0.10}$ \\
2E1739.1-1210 & $7.55_{-0.19}^{+0.22}$ & $25.11_{-3.00}^{+2.11}$ & $1.46_{-0.37}^{+0.51}$ &$45.22_{-14.16}^{+16.42}$ & $10.83_{-0.04}^{+0.03}$ & $10.46_{-0.14}^{+0.07}$ & $10.46_{-0.14}^{+0.07}$ & $0.58_{-0.11}^{+0.12}$ \\
2MASSJ07594181-3843560 & $<6.58$ & ... & $1.08_{-0.46}^{+0.51}$ &$39.20_{-5.56}^{+7.93}$ & $<10.63$ & $<9.63$ & $>10.59$ & $>0.90$ \\
2MASSJ17485512-3254521 & $<6.14$ & ... & $1.45_{-0.47}^{+0.56}$ &$44.52_{-16.37}^{+18.07}$ & $<9.42$ & $<9.05$ & $>8.99$ & $>0.51$ \\
2MASXJ00253292+6821442 & $6.16_{-0.18}^{+0.39}$ & $25.34_{-5.06}^{+2.79}$ & $1.46_{-0.40}^{+0.53}$ &$45.40_{-15.93}^{+16.38}$ & $9.63_{-0.05}^{+0.04}$ & $9.11_{-0.18}^{+0.10}$ & $9.11_{-0.18}^{+0.10}$ & $0.70_{-0.11}^{+0.11}$ \\
2MASXJ01064523+0638015 & $6.81_{-0.44}^{+0.69}$ & $21.56_{-7.50}^{+9.09}$ & $1.86_{-0.42}^{+0.53}$ &$44.58_{-9.10}^{+10.57}$ & $10.47_{-0.05}^{+0.04}$ & $<9.95$ & $>10.24$ & $>0.66$ \\
2MASXJ01073963-1139117 & $7.66_{-0.06}^{+0.09}$ & $25.52_{-1.96}^{+1.27}$ & $2.05_{-0.49}^{+0.58}$ &$41.53_{-10.35}^{+20.19}$ & $10.87_{-0.03}^{+0.03}$ & $10.62_{-0.10}^{+0.07}$ & $10.62_{-0.10}^{+0.07}$ & $0.44_{-0.12}^{+0.13}$ \\
2MASXJ03305218+0538253 & $6.76_{-0.33}^{+0.69}$ & $28.20_{-8.76}^{+6.57}$ & $2.46_{-0.61}^{+0.78}$ &$33.77_{-9.67}^{+8.47}$ & $10.81_{-0.06}^{+0.06}$ & $9.99_{-0.28}^{+0.23}$ & $9.99_{-0.28}^{+0.23}$ & $0.85_{-0.13}^{+0.10}$ \\
2MASXJ03342453-1513402 & $7.43_{-0.04}^{+0.05}$ & $26.61_{-1.09}^{+0.61}$ & $1.69_{-0.44}^{+0.58}$ &$43.00_{-16.10}^{+24.14}$ & $10.59_{-0.03}^{+0.03}$ & $10.51_{-0.05}^{+0.03}$ & $10.51_{-0.05}^{+0.03}$ & $0.18_{-0.10}^{+0.13}$ \\\bottomrule
\end{tabular}
\begin{tablenotes}
\item Notes -- \textit{Column 1:} Name of the source. \textit{Column 2:} Log of the dust mass in solar units. \textit{Column 3:} Dust temperature for the MBB component. '...' indicates the dust temperature was fixed at 23 K. \textit{Column 4:} Slope of the powerlaw component. \textit{Column 5:} Turnover wavelength. \textit{Column 6:} Log of the total infrared luminosity from 8--1000 \micron{} in solar units. \textit{Column 7:} Log of the MBB component luminosity in solar units. \textit{Column 8:} Log of the powerlaw component luminosity in solar units. \textit{Column 9:} Fractional contribution of the AGN to the total infrared luminosity calculated using Equations~\ref{eq:c12_lagnIR}~and~\ref{eq:c12_fagn}. The full version of this table is available in the online publication.
\end{tablenotes}
\end{threeparttable}
\end{table*}

Many models and templates exist in the literature to fit the broadband SEDs of galaxies. We chose a model that allowed us to both decompose the SED into star-forming and AGN components, as well as provide estimates on the dust temperature and dust mass.

\subsection{Casey 2012 Model}
One of the most widely used models for fitting the FIR SED of galaxies is a single modified blackbody (MBB). The simple model consists of a normal, single temperature blackbody that represents isotropic dust emission combined with a frequency dependent opacity given that dust is not a perfect blackbody. In the optically thin limit, the opacity can be approximated as a powerlaw, $\tau_{\nu}=(\nu/\nu_{0})^{\beta}$. The form of the single modified blackbody for the flux density at each frequency is then

\begin{equation}
S(\nu) \propto \nu^{\beta}B_{\nu}(T_{\mathrm{d}})\,,
\end{equation}

\noindent where $\beta$ is the spectral emissivity index and $B_{\nu}(T_{\mathrm{d}})$ is the standard Planck blackbody function for an object with temperature $T_{\mathrm{d}}$. This simple model has been shown to fit well the prominent FIR bumps for large samples of star-forming galaxies  and provides estimates of the dust temperature, dust mass, and SFR \cite[e.g.][]{Calzetti:2000fk, Bianchi:2013jk, Cortese:2014qq}.

To calculate the dust mass, we must assume a particular dust absorption coefficient, $\kappa_{0}$ at a particular frequency, $\nu_{0}$. For this work, we assume $\kappa_{0}=0.192\,\mathrm{m^{2}\,kg^{-1}}$ and $\nu_{0}=857\,\mathrm{GHz}$ (i.e. 350 \micron) from \citet{Draine:2003gd}. However, as \citet{Bianchi:2013jk} shows, by assuming a specific $\kappa_{0}$, we must also fix the spectral emissivity index to the value used to measure $\kappa_{0}$. In this work, we fix $\beta=2.0$ to match the spectral emissivity index used by \citet{Draine:2003gd}.\footnote{We have tested the effect of allowing $\beta$ to be a free parameter by re-fitting sources, which were detected in all wavebands. We find a median $\beta=1.8\pm0.3$, consistent with our choice to fix $\beta=2.0$. Further we find that all of the derived parameters, especially the integrated luminosities, are consistent within measurement uncertainty with their values when $\beta$ is fixed at 2.0.} The final full form of the single MBB model is then

\begin{equation}\label{eq:greybody}
S_{\mathrm{MBB}}(\nu) = \frac{M_{\mathrm{d}}\kappa_{0}}{D_{\mathrm{L}}^2}\left(\frac{\nu}{\nu_{0}}\right)^{\beta}\frac{2h\nu^{3}}{c^{2}}\frac{1}{e^{{h\nu/kT_{\mathrm{d}}}}-1}\,,
\end{equation}

\noindent where $M_{\mathrm{d}}$ is the dust mass, $D_{\mathrm{L}}$ is the luminosity distance, $c$ is the speed of light, $h$ is the Planck constant, and $k$ is the Boltzmann constant. The two free parameters then are $M_{\mathrm{d}}$ and $T_{\mathrm{d}}$, the dust mass and dust temperature respectively.

The simple assumption that dust emission in the IR can be modeled with a single temperature blackbody works well for ``normal'' star-forming galaxies. However, for galaxies with large amounts of hot dust either due to a compact starburst or central AGN, this assumption can quickly break down. To account for this hot dust we also fit our sample using the model described in \citet[][hereafter C12]{Casey:2012jl}, which is the combination of a single MBB and an exponentially cutoff powerlaw. The C12 model takes the form

\begin{equation}\label{eq:casey}
S_{\mathrm{C12}}(\nu) = N_{\mathrm{pl}}\left(\frac{\nu}{\nu_{\mathrm{c}}}\right)^{-\alpha}e^{-\nu_{\mathrm{c}}/\nu} + S_{\mathrm{MBB}}(\nu)\,,
\end{equation}

\noindent where $\nu_{\mathrm{c}}$ represents the turnover frequency and $N_{\mathrm{PL}}$ is a normalization constant. C12 illustrated using the \textit{Great Origins All-Sky LIRG Survey} (GOALS) sample that this model provides better estimates of the cold dust temperature, dust mass, and IR luminosity compared to both a single temperature modified blackbody and template libraries.

The C12 model introduces three more free parameters ($N_{\mathrm{PL}}$, $\alpha$, and $\nu_{\mathrm{c}}$); however, within the implementation used by C12, $N_{\mathrm{PL}}$ and $\nu_{\mathrm{c}}$ are tied to the normalization of the modified blackbody component and dust temperature to produce a smoothly varying SED and reduce the number of free parameters from five to three. 

But, after early tests using this initial setup, we found that fixing $N_{\mathrm{PL}}$ and $\nu_{\mathrm{c}}$ as a function of the other parameters produced unreliable fits. This is because AGN SEDs from the MIR to FIR are not as smooth as those seen in (U)LIRGS, likely due to the disconnect between star-formation and AGN heating. Within starbursting galaxies both the hot and cold dust are related through the same heating process, i.e. star formation, while much of the MIR emission in AGN host galaxies originates from dust around the AGN with no strong connection to global star formation in the galaxy. Therefore, we chose to leave both $N_{\mathrm{PL}}$ and $\nu_{\mathrm{c}}$ as free parameters resulting in a total of five for the entire model.

We describe in detail within Appendix~\ref{sec:sed_bayes} the exact fitting method used to estimate each parameter. In short we use a Bayesian framework with a Monte Carlo Markov Chain (MCMC) to probe the posterior probability distribution function (PDF). We marginalize the resulting posterior PDF for each parameter and use the median as our best-fit value and the 16th and 84th percentiles to define our lower and upper uncertainties, respectively. 

For 35 sources in our sample, the SEDs contain less than four detected points. These sources are undetected for several reasons. Either they are a) intrinsically faint, b) at relatively large distances, or c) located in a region with high foreground cirrus emission. This last reason causes problems because it produces high upper limits on the fluxes for the SPIRE wavebands, which give the appearance that a substantial amount of FIR emission could exist. Thus, we tested our modeling of these sources with three fixed dust temperatures: 15, 23, and 40K. 15 and 40 K are the extreme low and high temperatures we find for the whole sample, while 23 K is the median dust temperature. We fit only the 70 -- 500 \micron{} points/upper limits and let the dust mass increase until the model SED exceeded one of the points/upper limits. 

We find that using a dust temperature of either 23 or 40 K produces consistent MBB luminosities. Using a 15 K dust temperature results in MBB luminosities \textit{lower} than the 23 K and 40 K ones. Thus, it is reasonable to assume conservative upper limits on the MBB luminosity can be constrained with a higher dust temperature. The parameter most affected by the choice of temperature is the dust mass. Lower temperatures produce higher upper limits on the dust mass. In fact, moving from a 15 K temperature to a 40 K temperature changes the dust mass by two orders of magnitude. Given these tests, we chose to run our modeling of these 35 sources by fixing the dust temperature at 23 K, the median temperature of our whole sample. We acknowledge that for some of the undetected sources, we might be underestimating the upper limit on the dust mass. Since this subsample is only $\sim10$ per cent of the whole, we do not anticipate any large effect on the overall results.

We fit all 313 of the \herschel-BAT AGN using the modified C12 model and report the best fitting parameters in Table~\ref{tab:c12_params}. Best-fit model SEDs are shown along with the observed SEDs in Figure~\ref{fig:seds}. We also fit HRS galaxies using the exact same model so we can accurately compare the star-forming properties between non-AGN and AGN host galaxies. For the HRS sources with less than four detected points we fix the dust temperature at 21 K, the median dust temperature for the whole HRS sample. 

\subsection{Luminosities and AGN Fractions}\label{sec:lums_agn_frac}
In addition to the parameters associated with the model, we also calculated several luminosities and an AGN fraction. For this work, we define three luminosities: 1) $L_{\rm IR}$ will represent the total infrared luminosity determined by integrating the full SED from 8--1000 \micron, 2) $L_{\rm SF}$ will represent the 8--1000 \micron{} luminosity due to star formation in the host galaxy, 3) $L_{\rm AGN, IR}$ will represent the 8--1000 \micron{} luminosity due to AGN-heated dust. 

To determine $L_{\rm SF}$ and $L_{\rm AGN, IR}$ we first start with $L_{\rm PL}$ and $L_{\rm MBB}$, the 8--1000 \micron{} luminosities of the separate PL and MBB components from our SED decomposition. If we made the assumption that the PL component is completely dominated by the AGN in all sources then we could simply represent $L_{\rm PL}$ as the $L_{\rm AGN, IR}$ and $L_{\rm MBB}$ as $L_{\rm SF}$. However, we quickly recognized the well known effect that the PL component for some AGN host galaxies can have a strong contribution from dust heated by stars. Thus, a correction factor must be applied to convert $L_{\rm PL}$ and $L_{\rm MBB}$ into $L_{\rm AGN, IR}$ and $L_{\rm SF}$.

\begin{figure}
\includegraphics[width=\columnwidth]{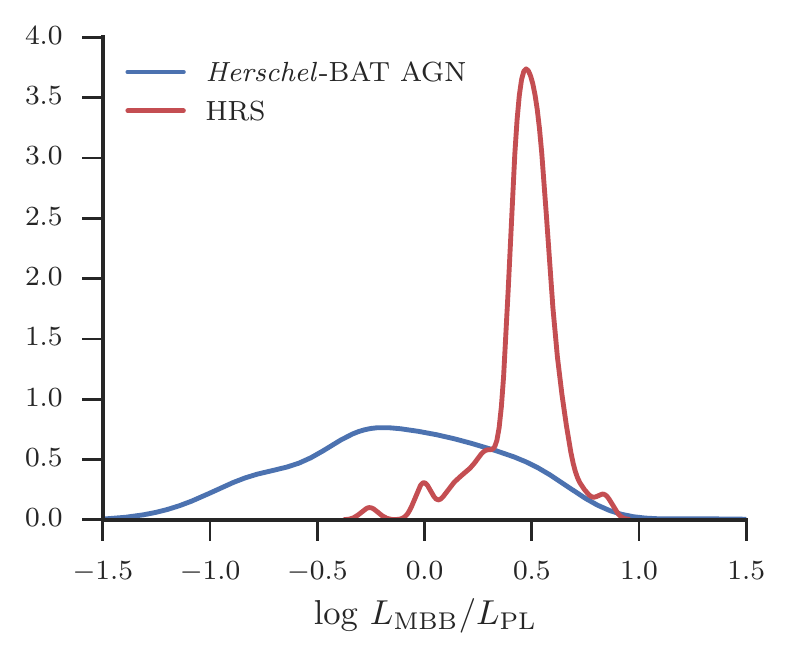}
\caption{KDEs of the $L_{\rm MBB}/L_{\rm PL}$ distribution for the \herschel-BAT AGN (blue), and HRS (red) samples. The HRS galaxies have a narrowly distributed ratio, while the BAT AGN span a wide range due to the AGN contribution. A color version of this figure is available in the online publication. \label{fig:lmbb_lpl_ratio}}
\end{figure}

We make this correction using the results of the C12 modeling for the HRS sample. Since all of the galaxies in this sample have either no or low-luminosity AGN, the MIR emission is primarily the result of stochastically heated grains near star-forming regions. In Figure~\ref{fig:lmbb_lpl_ratio}, we show the distribution of $L_{\rm MBB}/L_{\rm PL}$, the ratio of the MBB component luminosity to the MIR powerlaw component. The ratio for the HRS sample is narrowly distributed around a single value, indicating that for our lower mass non-AGN comparison sample the energy contained in the PL and MBB components are tightly connected. We find a median $\log(L_{\rm MBB}/L_{\rm PL})=0.48 \pm 0.13$, which transforms to $L_{\rm PL,SF} \approx \frac{1}{3}L_{\rm MBB}$, where we now indicate the contribution to the PL component from star formation as  $L_{\rm PL,SF}$. We can calculate the AGN contribution for the \herschel-BAT AGN then by assuming the star-forming emission follows the same ratio.

\begin{align}
L_{\rm AGN,\,IR} &= L_{\rm PL} - L_{\rm PL, SF} = L_{\rm PL} - \frac{1}{3}L_{\rm MBB} \label{eq:c12_lagnIR}\,, \\
f_{\rm AGN} &= L_{\rm AGN, \, IR}/L_{\rm IR} \label{eq:c12_fagn}\,, \\
L_{\rm SF} &= L_{\rm MBB} + L_{\rm PL,SF} =  \frac{4}{3}L_{\rm MBB} \label{eq:c12_lsf}\,,
\end{align}

\begin{figure}
\includegraphics[width=\columnwidth]{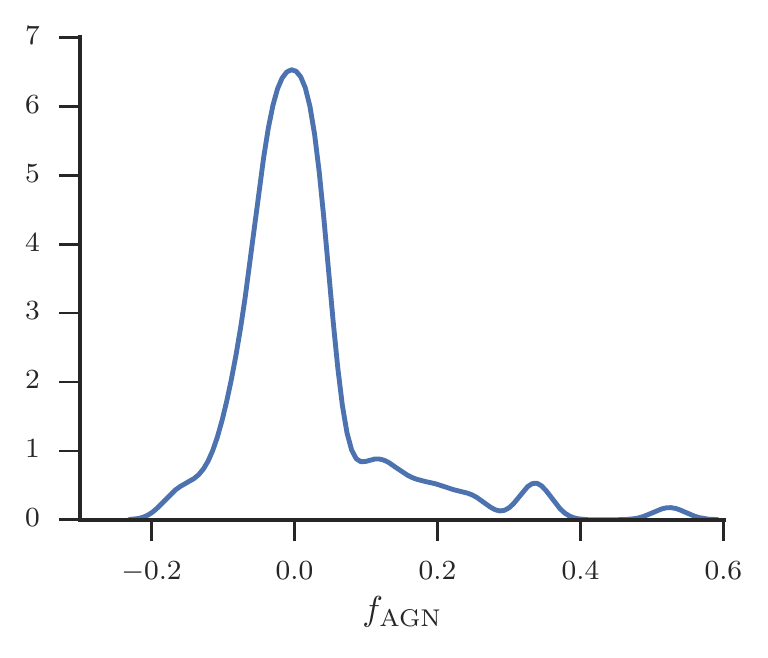}
\caption{$f_{\rm AGN}$ distribution for HRS. The standard deviation of this distribution quantifies the uncertainty on $f_{\rm AGN}$ associated with the correction factor used to calculate $f_{\rm AGN}$. A color version of this figure is available in the online publication. \label{fig:fagn_nonAGN}}
\end{figure}

The uncertainty on the correction factor leads to an uncertainty on $f_{\rm AGN}$ using this method, which we estimate by measuring $f_{\rm AGN}$ for the HRS sample. Figure~\ref{fig:fagn_nonAGN} shows the distribution of ``$f_{\rm AGN}$'' for HRS. As expected the distribution is centered around 0. We estimate the spread of the distribution by calculating the standard deviation, finding an uncertainty of 0.1. Therefore, we add in quadrature an uncertainty of 0.1 to all of our $f_{\rm AGN}$ estimates. Any $f_{\rm AGN}$ below 0.1 are converted to upper limits with a value of 0.1. 

In Appendix~\ref{sec:comp_models}, we detail our comparison between our chosen C12 model and two other template-based models that aim to decompose the IR SEDs of AGN host galaxies. We find broad agreement between all three models in the measurement of these three luminosities and the AGN fraction. These comparisons also provided additional systematic uncertainties on these parameters associated with model choice that we apply for the rest of this work. Further in Appendix~\ref{sec:pah_comp}, we analyze the relationship between our measured $L_{\rm SF}$ and polycyclic aromatic hydrocarbon (PAH) luminosities from \textit{Spitzer}/IRS spectra and compare the relationship with a star-forming galaxy sample. We determine that the relationship between our measured $L_{\rm SF}$ and PAH luminosities is consistent with that seen in a purely star-forming sample.

As explained in Appendix~\ref{sec:sed_bayes}, we used flat priors for all of the parameters in the C12 model except for $\nu_{c}$, the turnover frequency, for which we used a Gaussian prior centered at 45 \micron{} and a spread of 20 \micron. We imposed this prior due to how poorly $\nu_{c}$ is constrained in individual sources. Therefore, if the ``intrinsic'' AGN SED in reality extends to longer wavelengths than presumed, it is possible we are underestimating $L_{\rm AGN,\,IR}$ for our sources and overestimating $L_{\rm SF}$. \citet{Symeonidis:2016fk} recently showed for powerful AGN ($L_{X} > 10^{43.5}$ ergs s$^{-1}$), AGN heated dust might contribute more than expected in the FIR. The \citet{Symeonidis:2016fk} intrinsic AGN SED falls much more slowly with increasing wavelength than our model and other previous templates for the AGN SED (see their Figure 12). 

We test the effect of the \citet{Symeonidis:2016fk} SED by replacing the mean AGN SED used in our implementation of the DecompIR model with the \citet{Symeonidis:2016fk} SED since we show in Appendix~\ref{sec:comp_models} that DecompIR and C12 broadly agree in the determination of the various luminosities. We only fit the sources with more than 4 detected points in the SED to ensure the SED is sampled well. We calculate a median difference of only 0.01 dex between the $L_{\rm AGN,\,IR}$ using the mean AGN SED from DecompIR and the $L_{\rm AGN,\,IR}$ using the \citet{Symeonidis:2016fk}. The standard deviation of the difference is 0.2 dex, consistent with the uncertainties we find in Appendix~\ref{sec:comp_models}. Therefore, we conclude that the uncertainties on our estimates of the luminosities and AGN fractions adequately encompass the possibility of the AGN SED extending farther into the FIR than expected.

\section{Results and Discussion}

\subsection{Correlation between $L_{\rm AGN, IR}$ and $L_{\rm 14-195\,keV}$}\label{sec:lagnIR_lbat_correlation}

\begin{figure}
\includegraphics{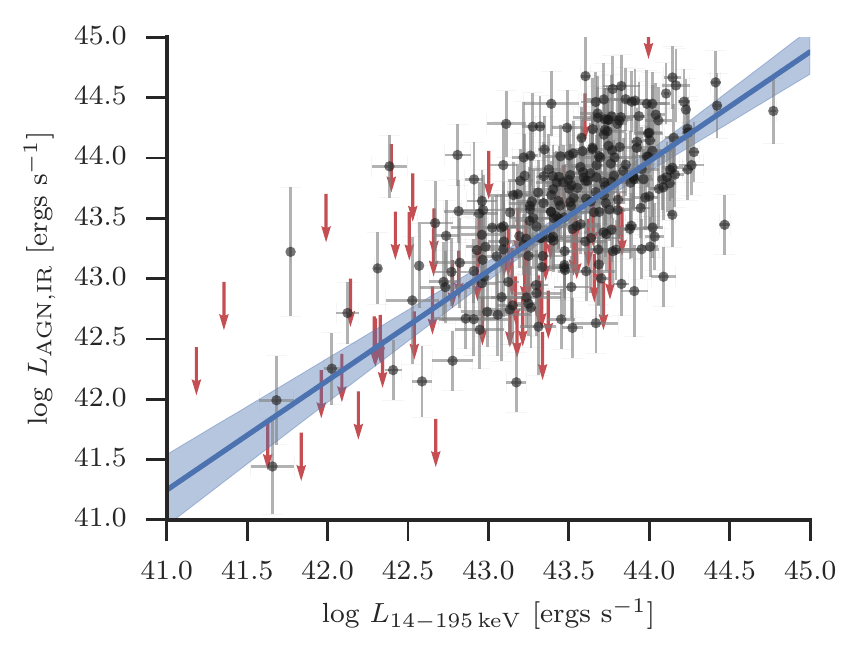}
\caption{Correlation between $L_{\rm AGN, IR}$ and $L_{\rm 14-195\,keV}$ for the \herschel-BAT sample (black points). The solid blue line shows the result of Bayesian linear regression with the shaded region indicating the 95 per cent confidence interval. The strong correlation between the two luminosities provides evidence our SED decomposition performed well in measuring the IR luminosity associated with AGN heated dust.\label{fig:lagnIR_lbat}}
\end{figure}

In Figure~\ref{fig:lagnIR_lbat}, we plot the relationship between the 14--195 keV luminosity and $L_{\rm AGN,IR}$. As expected, we find a strong correlation between the two luminosities indicating we have measured $L_{\rm AGN,IR}$ accurately from the C12 model and our correction factor. To quantify the relationship, we performed linear regression using a \textsc{python} implementation of \textsc{linmix\_err}\footnote{\url{https://github.com/jmeyers314/linmix}} \citep{Kelly:2007lr}, a Bayesian method for linear regression that incorporates errors in both independent and dependent variables, as well as non-detections. \citet{Kelly:2007lr} showed through simulations that \textsc{linmix\_err} outperforms other popular linear regression methods such as ordinary least squares and \textsc{fitexy}. \textsc{linmix\_err} outputs marginalized posterior probability distributions for the slope, intercept, and intrinsic scatter between the two variables, as well as the correlation coefficient, $\rho$. Our reported best fit parameters here and throughout the rest of the Paper will be the median of the marginalized posterior probability distribution with an uncertainty equal to the standard deviation. The linear model assumed is of the form:

\begin{equation}\label{eqn:lin_mod}
Y = mX + b + \epsilon_{\rm int}\,,
\end{equation}

\noindent where $Y$ and $X$ are the dependent and independent variables, $m$ is the slope, $b$ is intercept, and $\epsilon_{\rm int}$ represents intrinsic scatter in the relationship beyond that related to measurement error. $\epsilon_{\rm int}$ is modeled as a Gaussian random variable with mean 0 and a variance of $\sigma^2$. 

We plot the best fit linear regression for the whole sample in Figure~\ref{fig:lagnIR_lbat} as a blue line with shading to indicate the 95 per cent confidence interval. The best-fit line corresponds to the following relationship:

\begin{equation}\label{eqn:lagnIR_lbat}
\log(L_{\rm AGN,IR}) =  (0.91\pm0.06)\log(L_{\rm 14-195\,keV}) - (4.03\pm2.6)\,,
\end{equation}

The correlation coefficient from \textsc{linmix\_err} is $0.8\pm0.03$ and the intrinsic scatter is 0.15 dex. The probability of zero correlation is only $10^{-77}$. Our measured slope of 0.91 agrees perfectly with the measured slope of $0.91\pm0.04$ in \citet{Asmus:2015qy} using the nuclear 12 \micron{} emission from high angular resolution imaging. Given our very simple method for determining the AGN MIR luminosity from broadband SED decomposition, a slope consistent with the more robust \citet{Asmus:2015qy} relationship and high correlation coefficient provides confidence that the decomposition has performed reasonably well. Further, Equation~\ref{eqn:lagnIR_lbat} will provide future studies with estimates of the full 8--1000 \micron{} AGN contribution and allow better estimates of IR-based SFRs for AGN, as long as X-ray data are available.

Because we are comparing luminosities and the \swift/BAT sample is flux-limited, the correlation we see could simply be due to a confounding variable, in this case distance. To investigate this possibility, we performed a partial correlation test with ASURV using distance as our third variable along with the 14--195 keV luminosity and AGN-related IR luminosity. The partial correlation test calculates Kendall's-$\tau$ rank correlation coefficient between all three variables, then estimates the intrinsic correlation coefficient after removing the dependence on the third variable (i.e. distance). 

We find $\tau = 0.30\pm0.03$ and reject the null hypothesis of zero partial correlation at a significance level of $10^{-23}$. While the value of the correlation coefficient is less than the correlation coefficient from \textsc{linmix\_err}, the partial correlation test does not account for scatter due to measurement error. 

To further show that our SED decomposition has accurately estimated the IR luminosity due to the AGN, we calculated the correlation between just the total 8--1000 \micron{} luminosity ($L_{\rm IR}$) and 14--195 keV luminosity using the same method. We find a slope of $0.41\pm0.05$ and a correlation coefficient of only $0.44\pm.05$. A z test between the $L_{\rm AGN,\,IR}$ correlation coefficient and the $L_{\rm IR}$ correlation coefficient indicates a zero probability that the $L_{\rm IR}$-$L_{\rm X}$ correlation is stronger than the $L_{\rm AGN,\,IR}$-$L_{\rm X}$ correlation. 

14--195 keV luminosities are only available for our low redshift sample, therefore Equation~\ref{eqn:lagnIR_lbat} is useless at higher redshifts. We can convert 14--195 keV luminosity to the more accessible 2--10 keV luminosity by making the simple assumption that the shape of the X-ray spectrum of AGN is a powerlaw with a photon index of 1.8 \citep[e.g.][]{Vasudevan:2013dz}. The ratio of $L_{14-195}/L_{2-10}$ is then $\sim2.7$. Switching to $L_{2-10}$ changes Equation~\ref{eqn:lagnIR_lbat} to 

\begin{equation}\label{eqn:lagnIR_2_10}
\log(L_{\rm AGN,IR}) =  (0.91\pm0.06)\log(L_{\rm 2-10\,keV}) - (3.64\pm2.6)
\end{equation}

\noindent $L_{\rm 2-10\,keV}$ assumes the luminosity has been absorption-corrected. Differences in the assumed photon index only change the intercept by at most 0.3 dex. 

\subsection{IR colors as a predictor of $f_{\rm AGN}$ and selecting AGN}

\begin{figure*}
\includegraphics[width=\textwidth]{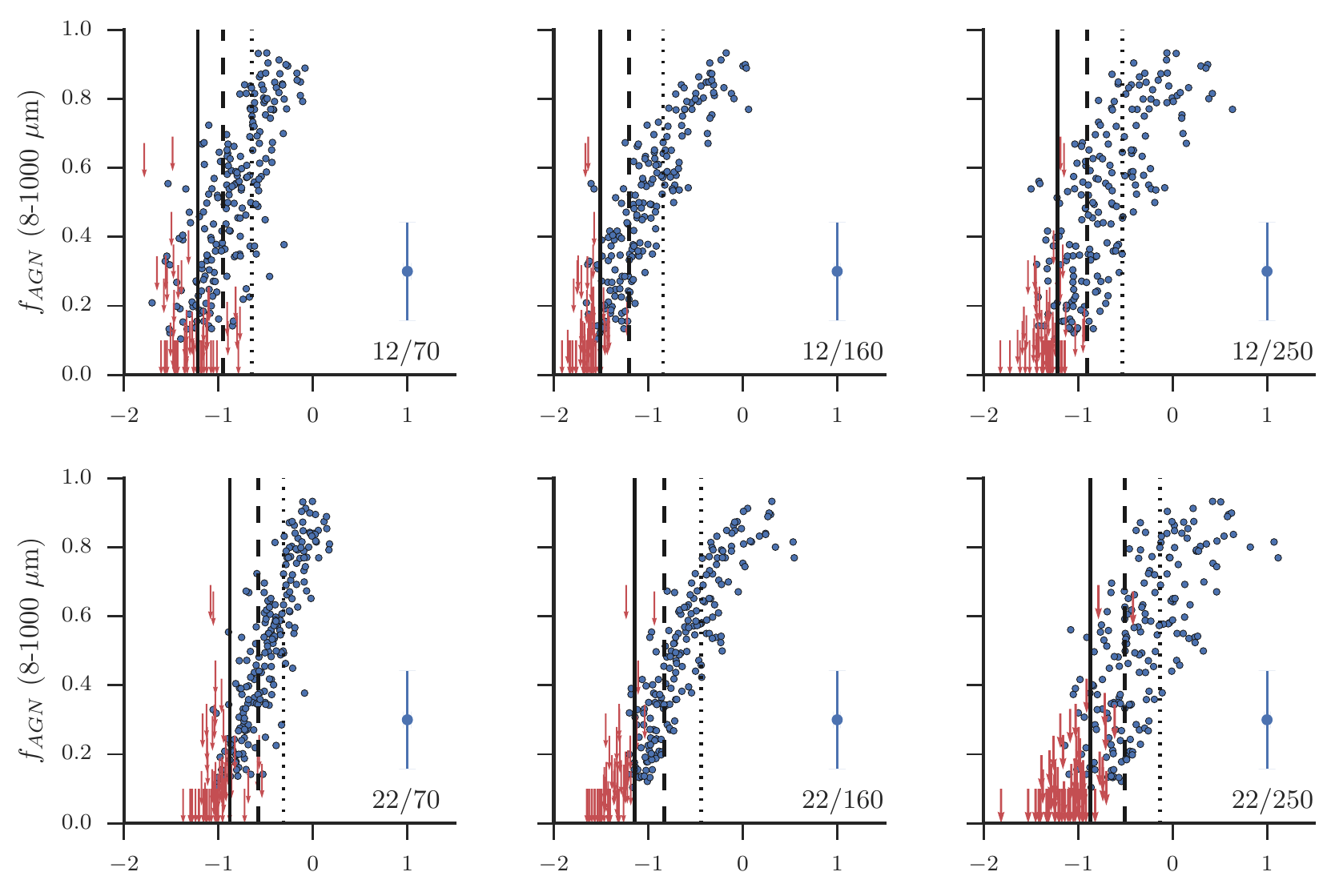}
\caption{Relationship between different IR colors and $f_{\rm AGN}$, the fraction of the 8--1000 \micron{} luminosity attributed to AGN heated dust. The IR colors (x-axis) are in logarithmic units. Upper limits on $f_{\rm AGN}$ are indicated as red arrows and are at 95 per cent confidence. We also display the 75th, 50th, and 25th percentile colors for the \herschel-BAT AGN as solid, dashed, and dotted vertical lines, respectively. Choosing a sample with colors \textit{above} these would result in 75, 50, and 25 per cent completeness for the \herschel-BAT AGN. \label{fig:agnfrac_vs_flux_ratio}}
\end{figure*}

We utilize our broad coverage of the IR SED for AGN host galaxies to test the viability of different IR colors for predicting the AGN IR luminosity fraction (i.e. $f_{\rm AGN}$). For large surveys and/or high redshift studies, coverage from 12--500 \micron{} might not be available or possible.

In Figure~\ref{fig:agnfrac_vs_flux_ratio}, we plot six MIR/FIR flux ratios (12/70, 12/160, 12/250, 22/70, 22/160, 22/250) against the measured $f_{\rm AGN}$ for the \herschel-BAT AGN. The flux ratios are all in log units to better visualize the relationships. Upper limits on $f_{\rm AGN}$ are shown as downward-pointing arrows. In the lower right corner of each subplot, we show representative error bars. We fit a linear model between each flux ratio and $f_{\rm AGN}$ with the following form

\begin{equation}
f_{\rm AGN} = m\log(F_{1}/F_{2}) + b + \epsilon_{\rm int}\,,
\end{equation}

\noindent where $F_{1}/F_{2}$ represents each flux ratio. To calculate the best fitting parameters, we again used \textsc{linmix\_err}. 

Table~\ref{tab:flux_ratio_fagn_linreg} outlines the best fit parameters for each relationship. Each flux ratio is highly correlated with $f_{\rm AGN}$ (all $\rho > 0.87$). Figure~\ref{fig:avg_sed} also reemphasizes this result as the MIR fluxes change more than the FIR ones as the X-ray luminosity changes. The worst indicators are the flux ratios involving the 250 \micron{} flux with $\rho$ less than 0.9 and significant scatter as shown in Figure~\ref{fig:agnfrac_vs_flux_ratio}. The 250 \micron{} flux is located on the Rayleigh Jeans side of the blackbody and varies much more strongly as a function of the dust temperature due to the rapidly declining shape of the SED at long wavelengths compared to the 160 \micron{} flux, which is nearer to the peak of the MBB component. Thus, the 160 \micron{} flux is more tightly related to the total MBB luminosity and $f_{\rm AGN}$.

Based on the slope of the linear regressions, the 22/70 \micron{}, with a slope of 0.8, overall seems to be the best flux ratio to use as an indicator of the AGN contribution to the IR SED. These two wavebands bracket the region where the intrinsic AGN SED is thought to turnover \citep[e.g.][]{Netzer:2007ve, Mullaney:2011yq} and seems to be the crucial transition region where the SED shifts from AGN-dominated to host galaxy dominated. The 22/70 \micron{} color is also quite similar to the long used 25/60 \micron{} color derived from the IRAS survey as an indicator of AGN activity \citep{de-Grijp:1985lr,de-Grijp:1987gb}. 

\begin{table*}
\begin{threeparttable}
\captionsetup{font=small,labelfont=bf,labelsep=period}
\caption{Linear Regression Between Flux Ratios and $f_{\rm AGN}$\label{tab:flux_ratio_fagn_linreg}}
\begin{tabular}{lccccccc}
\toprule 
$F_{1}/F_{2}$ & $m$ & $b$  & $\sigma_{\rm int}$ & $\rho$ & 75th \%tile  & 50th \%tile  & 25th \%tile\\
&&&&&(\%tile for HRS)&(\%tile for HRS)&(\%tile for HRS)\\
\midrule
12/70      & $0.67\pm0.04$ & $1.02\pm0.03$ & $0.016\pm0.004$ & $0.89\pm0.02$ & -1.23 (90) & -0.97 (50) & -0.66 (21)\\
12/160    & $0.58\pm0.02$ & $1.05\pm0.03$ & $0.002\pm0.002$ & $0.99\pm0.01$ & -1.51 (31) & -1.21 (5.2) & -0.85 (2.2)\\
12/250    & $0.51\pm0.03$ & $0.84\pm0.03$ & $0.019\pm0.004$ & $0.87\pm0.03$ & -1.22 (28) &  -0.91 (5.2) & -0.57 (1.5)\\
22/70      & $0.79\pm0.03$ & $0.86\pm0.02$ & $0.001\pm0.001$ & $0.99\pm0.01$ & -0.89 (32) & -0.60 (16) & -0.32 (12)\\
22/160    & $0.58\pm0.02$ & $0.85\pm0.02$ & $0.0004\pm0.0005$ & $0.997\pm0.003$ & -1.14 (5.6) & 0.83 (1.2) & -0.45 (0)\\
22/250    & $0.49\pm0.03$ & $0.65\pm0.02$ & $0.016\pm0.004$ & $0.90\pm0.02$ & 0.87 (6.8) & -0.51 (0.02) & -0.18 (0) \\
\bottomrule
\end{tabular}
\end{threeparttable}
\end{table*}

In Figure~\ref{fig:agnfrac_vs_flux_ratio} we also plot the 75th, 50th, and 25th percentile colors for the \herschel-BAT sample as black solid (25th), dashed (50th), and dotted (75th) vertical lines. In Table~\ref{tab:flux_ratio_fagn_linreg}, the last three columns list the colors in logarithmic units for the 75th, 50th, and 25th percentiles. Thus, if these colors were used as a selection method; 75, 50, and 25 per cent of the \herschel-BAT sample would have been selected.

While these colors are useful for selecting AGN host galaxies, we also want to know what percentage of non-AGN galaxies would be selected with these color cuts as contamination for a sample. Therefore, we also calculated the percentile of HRS galaxies that would be selected. For this analysis, we used the complete HRS sample since it represents the natural field galaxy population, and we do not want to restrict to only high-mass non-AGN galaxies. The HRS sample, unfortunately, was not observed at 70 \micron, therefore to calculate the 12/70, and 22/70 colors, we simply used the 70 micron flux from the best-fit model. The percentiles for HRS are given in parentheses next to the values of the colors in Table~\ref{tab:flux_ratio_fagn_linreg}. 

We find that the best color to use in selecting an AGN-only sample would be the 22/160 color. Using the 75th percentile color (-1.14), we would only select 5.6 per cent of the HRS sample. To get only around 1 per cent of the HRS sample, we would have to use the 50th percentile 22/160 or 22/250 color. To select no HRS galaxies would require pushing to the 25th percentile colors and risk missing a large fraction of the \herschel-BAT AGN. We note this analysis is quite simple and does not take into account other contributing factors such as the flux-limit of \swift/BAT and how representative HRS is of the non-AGN population. Rather, we only mean to briefly explore how well certain colors would select our AGN and our comparison non-AGN. In a forthcoming paper, we aim to evaluate these colors along with other MIR color selection techniques \cite[e.g.][]{Donley:2012qy, Stern:2012mz} to determine the biases and selection effects with regards to the SFR of the galaxy and strength of the AGN.

\subsection{The AGN contribution to the 70 \micron{} emission}\label{sec:agn_70_contribution}
Many studies examining the relationship between nuclear activity and star formation rely on single monochromatic luminosities as an indicator of star formation \cite[e.g.][]{Netzer:2007ve, Netzer:2009lr, Rosario:2012fr}. The most widely used indicator has been the 60 \micron{} luminosity given its availability from the all-sky \textit{IRAS} survey. 

With our \herschel-BAT sample and SED modeling we can test how often and at what level AGN emission contributes to different wavelength bands. We test the AGN contribution to the 70 \micron{} emission. Based on visual inspection of all of the SED fits, we found this waveband is the only \herschel{} band with a noticeable AGN contribution. All other longer wavelength bands were dominated by the MBB component. We restrict this analysis to only those sources with well-determined SEDs longwards of 70 \micron, excluding those objects with less than four detected SED points. We note that the majority of these objects are likely AGN dominated at 70 \micron; however, the uncertainty in their FIR SED, especially their dust temperature, results in a largely unconstrained calculation.

To determine the AGN contribution to the 70 \micron{} emission, we first estimated the PL component associated with star formation. We assumed the shape of the star-forming PL component followed the average power law slope ($\alpha=0.0$ ) and turnover wavelength ($\lambda_{\rm c} = 50$ \micron) from the HRS sample. We then adjusted the PL normalization such that the integrated luminosity of the PL component would equal $1/3L_{\rm MBB}$, which is the ratio found for non-AGN galaxies from the HRS sample (see Section~\ref{sec:lums_agn_frac}). We combined the estimated star forming PL component with the measured MBB component to form each of the \herschel-BAT AGN's total star-forming component. The 70 \micron{} AGN contribution was then calculated as the excess 70 \micron{} emission leftover after subtracting the star-forming contribution from the total 70 \micron{} emission from the best-fit SED. We denote the fraction contributed by the AGN as $f_{\rm AGN, 70}$.

\begin{figure*}
\includegraphics{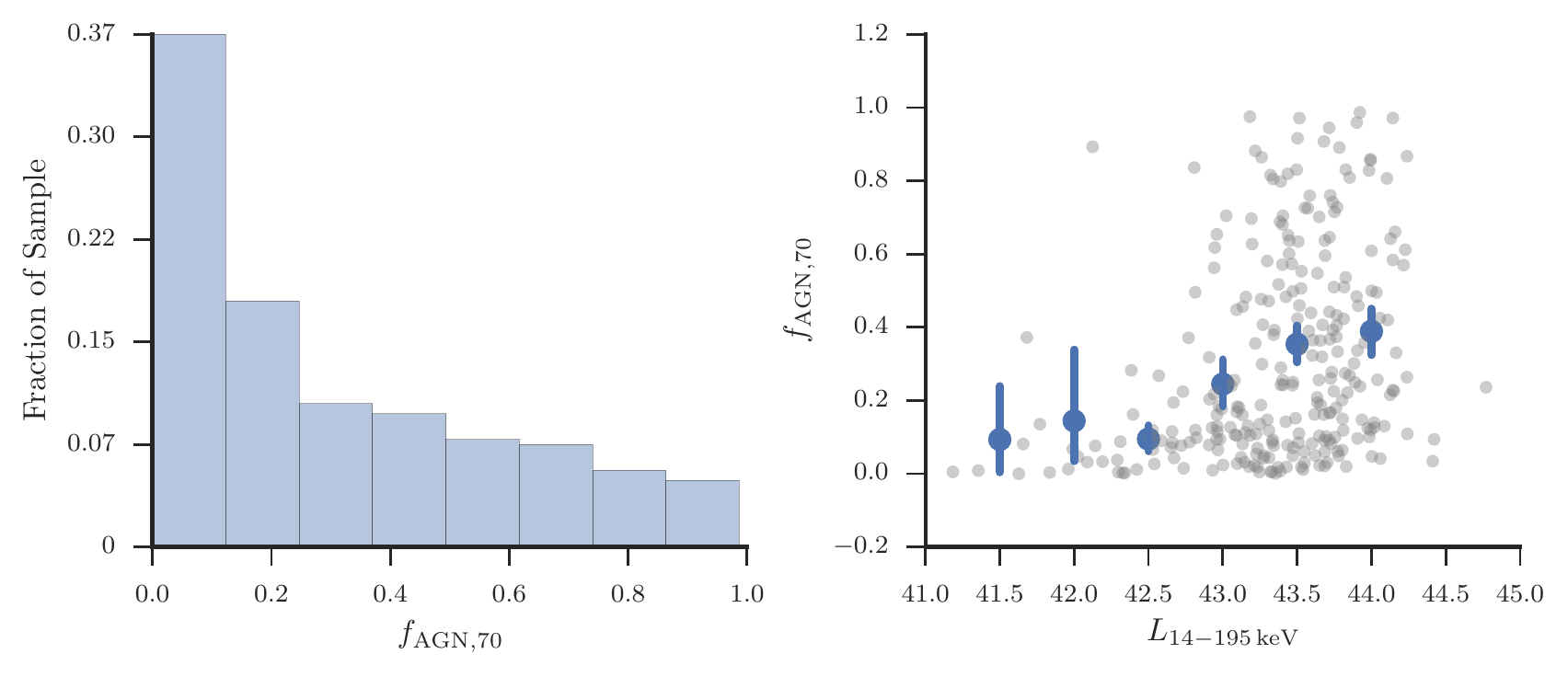}
\caption{\textit{left:} Histogram of the estimated 70 \micron{} contribution by AGN-heated dust. Each bar is a bin with a width of 0.1 and a height equal to the fraction of the total sample. \textit{right:} Relationship between the 14--195 keV luminosity and 70 \micron{} AGN contribution. Grey dots represent each individual source while the blue dots with error bars are averages within bins according to 14--195 keV luminosity. Error bars indicate the 68 percent confidence interval determined using bootstrap analysis.\label{fig:70um_agn_contribution}}
\end{figure*}

Figure~\ref{fig:70um_agn_contribution}, \textit{left} shows the distribution of $f_{\rm AGN, 70}$. We find that for 48 per cent of the sample,  $f_{\rm AGN, 70} < 0.2$ but for 24 per cent of the sample $f_{\rm AGN, 70} > 0.5$. Using the 70 \micron{} luminosity as a single SFR indicator would overestimate values by at least a factor of 2 for almost one-third of an AGN sample. 

In the right panel of Figure~\ref{fig:70um_agn_contribution} we show $f_{\rm AGN, 70}$ as a function of the 14--195 keV luminosity. The grey points indicate the individual points while the blue points with error bars are binned averages with a 68 per cent confidence interval determined through bootstrapping. It is clear that at $\log\,L_{\rm 14-195\,keV} < 42.5$, $f_{\rm AGN, 70}$ is largely negligible with only two sources with $f_{\rm AGN, 70} > 0.2$. These AGN are NGC 1052 and NGC 4941. Both are quite pointlike, possibly unresolved at 70 \micron, while at long wavelengths show large-scale extended emission.

Above $\log\,L_{\rm 14-195\,keV} = 42.5$, the AGN contribution at 70 \micron{} varies nearly uniformly between 0 and 1. This indicates that the AGN contribution is not a simple function of high AGN luminosity, rather it is due to the competing interplay between AGN emission and star-formation. This reinforces the need for SED decomposition to determine accurate SFRs when an AGN is present. We repeated this process at 160 \micron, finding no source with an AGN contribution above the 5 per cent level.

A possible complication relating to the calculation of the AGN contribution at 70 \micron{} is the effect of stochastically heated small grains. These grains are hotter than the relatively cooler large grains that produce the majority of the long wavelength FIR emission. In starburst galaxies, the small grains can provide a large amount emission at warmer dust temperatures which results in enhanced emission in the 30--70 \micron{} region and a steep MIR powerlaw slope. Indeed, \citet{Casey:2012jl} uses her model to fit the SEDs of local ULIRGs. However, none of the \herschel-BAT AGN would be classified as a ULIRG and very few are LIRGs. Further we showed in \citet{Shimizu:2015xo} that only a small percentage of our AGN are above the main sequence where starbursts are found. Therefore, we do not consider a strong starburst component to be affecting a large number of our sources and affecting our 70 \micron{} AGN contributions.
 
\subsection{$M_{\rm dust}$, $T_{\rm dust}$, and SFR of local AGN host galaxies}\label{sec:agn_sf_comparison}

\begin{figure*}
\includegraphics[width=\textwidth]{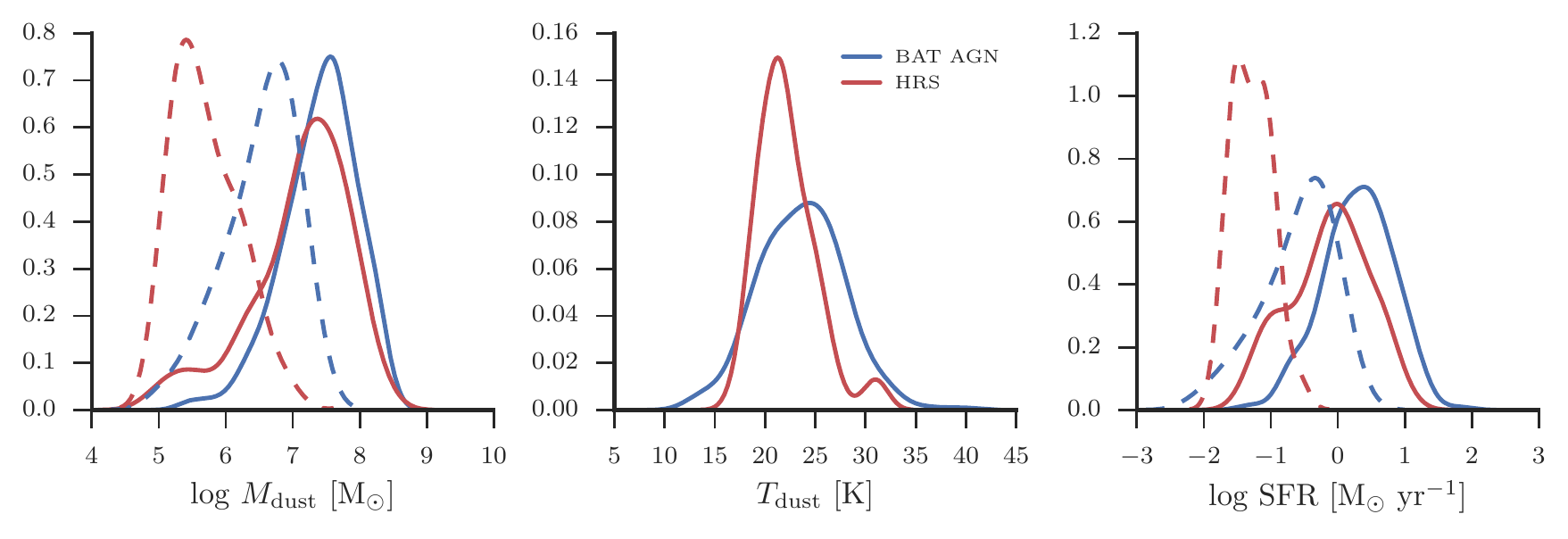}
\caption{ Comparison of $M_{\rm dust}$, $T_{\rm dust}$, and SFR between the \herschel-BAT AGN (blue), and HRS mass-matched sample (red). The distributions with dashed lines indicate sources with only upper limits on these values. A color version of this figure is available in the online publication. \label{fig:mdust_tdust_sfr_comp}}
\end{figure*}

\begin{table*}
\centering
\begin{threeparttable}
\captionsetup{font=small,labelfont=bf,labelsep=period}
\caption{Mean \mdust, \tdust, and SFR Including Non-Detections\label{tab:mean_mdust_tdust_sfr}}
\begin{tabular}{lcccccc}
\toprule 
Sample & $\log$ \mdust & \tdust & $\log$ SFR\\ 
 & [\msun]  & [K]  &  [\msun{} yr$^{-1}$] \\
\midrule
\herschel-BAT & 7.36 (6.56--7.90) & 23.8 (19.4--27.5) & 0.23 (-0.57--0.81)\\ 
HRS & 7.08 (5.12--7.70) & 21.5 (19.4--24.6) & -0.23 (-1.57--0.36) \\
\bottomrule
\end{tabular}
\end{threeparttable}
\end{table*}

Figure~\ref{fig:mdust_tdust_sfr_comp} compares the distribution of \mdust, \tdust, and SFR for our three samples of galaxies. We calculated SFRs using the conversion from 8--1000 \micron{} star-forming luminosity using the following equation from \citet{Murphy:2011rt}

\begin{equation}\label{eq:sfr_ir}
\rm{SFR} = \frac{L_{\rm SF}}{2.57\times10^{43}}\,,
\end{equation}

\noindent The solid lines show the KDE for sources which have firm measurements, while the dashed lines show the KDE for the sources with only upper limits. Table~\ref{tab:mean_mdust_tdust_sfr} displays the medians for these three properties and each sample. The values inside parentheses indicate the 16th and 84th percentiles for each distribution. The median, 16th, and 84th percentiles were calculated using survival analysis and the KMPL as in Section~\ref{sec:ir_seds} to account for upper limits. 

We find that the \herschel-BAT AGN have higher dust masses than the HRS sample, slightly higher dust temperatures, and higher SFRs. Because we have mass-matched the HRS sample to the \herschel-BAT sample, there should be no effects due to stellar mass. To confirm that the distributions are statistically different we run the Peto-Prentice two-sample test. The Peto-Prentice two sample test is comparable to the Kolmogorov-Smirnov (K-S) test, which calculates the probability that two separate samples were drawn from the same parent population. The Peto-Prentice test, however, allows for the inclusion of censored data. 

For the comparison of dust masses, we find the probability that the \herschel-BAT AGN and HRS galaxies are drawn from the same parent distribution is $1.8\times10^{-5}$. The probability for the dust temperatures is $4\times10^{-4}$, while the probability for the SFRs is $8.4\times10^{-11}$. All three probabilities are less than 0.05, the standard cutoff for declaring two samples are inconsistent with being drawn from the same parent population. 

The increased SFRs for the \herschel-BAT sample compared to the mass-selected HRS sample are consistent with the results of \citet{Koss:2011vn}. \citet{Koss:2011vn}, using optical colors, found that the \swift/BAT AGN displayed bluer colors compared to a mass-matched sample of Sloan Digital Sky Survey (SDSS) galaxies, indicative of increased levels of star formation. 

These results may seem contradictory with our previous work in \citet{Shimizu:2015xo} where, using the SFRs calculated here, we showed that AGN host galaxies are experiencing decreased levels of star formation given their stellar mass. However, the discrepancy lies in the comparison being done. In \citet{Shimizu:2015xo} we compare the \herschel-BAT AGN to the so-called star-forming ``main sequence,'' a tight nearly linear correlation between SFR and \mstar \citep{Brinchmann:2004bf, Salim:2007rm, Noeske:2007fr}, finding the AGN are falling off and transitioning towards quiescence. In this work, as well as in \citet{Koss:2011vn}, simple mass-matching is used to form comparison samples, which leads to selecting only high stellar mass non-AGN galaxies due to the high stellar masses of the BAT AGN. Based on previous studies of the color-magnitude relation, however, selection of high mass galaxies will result in a large quiescence fraction especially at low redshifts  \citep{Kauffmann:2003lr, Wuyts:2011lh, Bell:2012qf}. Therefore, since the \herschel-BAT AGN show both intermediate optical colors and SFRs, it makes sense that a mass-selected non-AGN sample will indicate bluer colors and higher SFRs for AGN host galaxies given quiescent galaxies are largely red with low SFRs.

Further, looking into the details of our mass-selected HRS sample, we find that 47/122 (50 per cent) are morphologically classified as early type (E/S0), the dominant morphology for quiescent galaxies \citep{Wuyts:2011lh, Bell:2012qf}. The \herschel-BAT AGN, though, are largely late type spirals \citep{Koss:2011vn}, and even the few ellipticals in our sample show higher SFRs than expected for quiescent galaxies \citep{Shimizu:2015xo}. Thus, we can reconcile the results of \citet{Koss:2011vn}, \citet{Shimizu:2015xo}, and this work by understanding that mass-matching preferentially selects low star-forming, red galaxies in contrast to the intermediate star-forming, blue AGN. All of this agrees with previous work comparing X-ray selected AGN with mass-matched non-AGN galaxies such as \citet{Santini:2012ly} who found enhanced SFRs in AGN host galaxies. This enhancement, however, disappeared after removing quenched galaxies from the comparison non-AGN sample.  

The increased dust temperatures observed in the \herschel-BAT AGN could be explained by the increased SFRs. Dust temperatures are known to increase with both infrared luminosity and specific SFR \citep[e.g][]{Chapman:2003jk, Symeonidis:2013fe, Cortese:2014qq, Magnelli:2014sf}, although these studies did not extend to the low luminosities and specific SFR we observe in our samples. Because we know our sample contains an AGN, it is possible at least some of the increased dust temperatures could be due to AGN heating. Recently, \citet{Garcio-Gonzalez:2016bk} spatially decomposed the FIR images of a small sample of nearby Seyfert galaxies and found that nuclear emission shows markedly higher dust temperatures than the host galaxy (as probed through the 70/160 \micron{} color). Sources that are dominated by AGN-related nuclear emission should then show higher temperatures that do not match with the overall correlation between SFR and $T_{\rm dust}$. 

\begin{figure*}
\includegraphics[width=\textwidth]{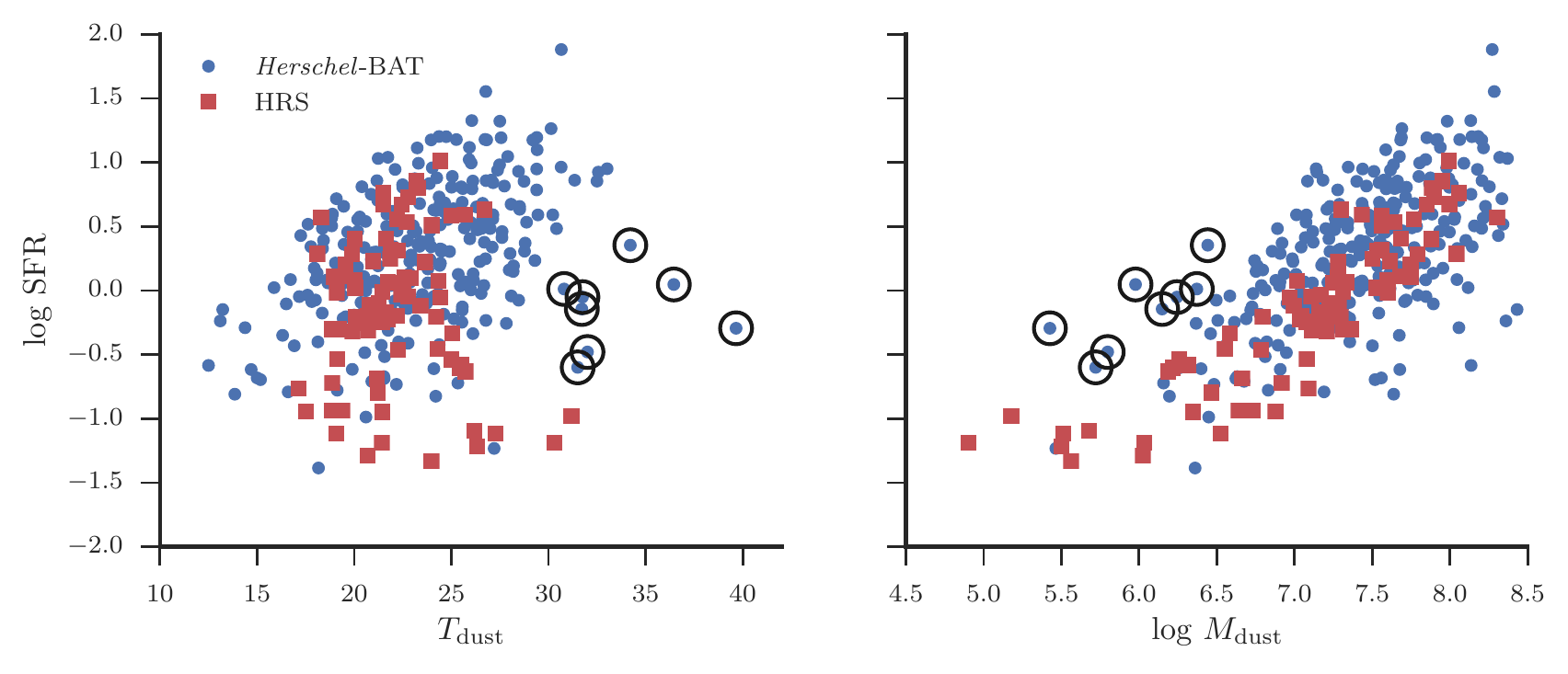}
\caption{\textit{left:} Correlation between dust temperature and SFR for the \herschel-BAT AGN (blue points) and HRS galaxies (red squares). Circled blue points indicate possible sources that are dominated by AGN-heated dust. \textit{right:} Correlation between dust mass and SFR for \herschel-BAT AGN and HRS galaxies. The circled points are again the sources which could be dominated by AGN-heated dust. \label{fig:sfr_tdust}}
\end{figure*}

In the left panel of Figure~\ref{fig:sfr_tdust}, we show the relationship between SFR and $T_{\rm dust}$ for the \herschel-BAT AGN and HRS galaxies. The bulk of both samples follow in general a broad positive correlation. There exists a group of HRS galaxies at low SFR ($\log\,\rm{SFR} < -1.0$) that show a flat relationship with $T_{\rm dust}$. All but one of these galaxies are early type, which are known to show dust temperatures up to 30 K \citep{Temi:2007ty}. The encircled \herschel-BAT AGN at $T_{\rm dust} > 30$ K and $\log\,\rm{SFR} < 0.5$ seem to represent distinct outliers from both samples. These sources are 2MASX J09235371-3141305, 2MASXJ 15462424+6929102, ESO 103-035, ESO 417-G006, MCG --05-23-016, MCG --06-30-015, Mrk 841, and NGC 3516. Except for 2MASX J09235371-3141305, these sources also all have $f_{\rm AGN} > 0.5$, indicating the MIR PL component is quite dominant. ESO 103-035, MCG --05-23-016, and Mrk 841 show SEDs that peak around 20 \micron, highly indicative of an SED dominated by IR emission from the AGN. For these seven AGN, then, it is possible that the large dust temperatures are due to AGN heating, rather than star formation.

Besides the seven sources listed, though, the bulk of the \herschel-BAT AGN seem to show \tdust{} largely consistent with star-formation heated dust. 

The higher SFRs could also explain the slightly larger dust masses observed for the \herschel-BAT AGN.\footnote{Excluding the objects with only upper limits on the dust mass, which are dependent on the assumed fixed dust temperature we used in the modeling (see Section~\ref{sec:sed_fit}), produces the same overall trend of higher dust masses in the \herschel-BAT AGN sample.} The \herschel-BAT AGN have a median $\log$\mdust{} of 7.38 \msun{} vs. 7.0 \msun{} for the HRS. The 16th percentiles also extend to much lower dust masses for the HRS. \citet{Cortese:2012xy} showed using the full HRS sample that the dust-to-stellar mass ratio is anti-correlated with $NUV-r$ color, a proxy for specific SFR and decreases strongly when moving from late type to early type galaxies. \citet{da-Cunha:2010qf} also showed, using over 3000 galaxies from SDSS, a strong correlation between the cold dust mass of galaxies and their SFR, likely a manifestation of the well known Schmidt-Kennicutt law \citep{Schmidt:1959fj,Kennicutt:1998wl} which links the SFR surface density with the surface density of gas.

In the right panel of Figure~\ref{fig:sfr_tdust}, we show the correlation between SFR and \mdust{} for both samples with the same colors and symbols as before. While both samples follow the same overall correlation, the HRS galaxies display a much tighter relationship between SFR and \mdust{} than the \herschel-BAT sample. The \herschel-BAT AGN seem to scatter more above the HRS galaxies such that for a fixed \mdust, the \herschel-BAT AGN have higher SFRs. If we assume that \mdust{} follows $M_{\rm gas}$, then on average the \herschel-BAT AGN would have higher SFR/$M_{\rm gas}$, a proxy for the star formation efficiency (SFE). This was tentatively seen in \citet{Maiolino:1997vn} and more recently in \citet{Saintonge:2012uq} with the COLDGASS sample where AGN showed decreased gas depletion times ($t_{\rm dep} = \rm{SFE}^{-1}$). Our result is also tentative given we are using the dust mass as a tracer of the gas mass, therefore CO observations of the \herschel-BAT AGN would be an intriguing study to test the results seen here and in \citet{Saintonge:2012uq}. Because of the tentative nature of this finding, we refrain from over-interpreting and simply note that the increased dust masses seen in Figure~\ref{fig:mdust_tdust_sfr_comp} can be explained by the increased SFRs of the \herschel-BAT AGN given the strong correlation between the two properties.

\subsection{The Correlation Between SFR and AGN Luminosity}
Our main goal in this Paper is to robustly assess the observational evidence for a connection between the AGN and star formation in its host galaxy in the local universe. To accomplish this, we analyze the correlation between the AGN strength and SFR. We assume the 14--195 keV luminosity ($L_{\rm 14-195 keV}$) linearly probes the bolometric AGN luminosity, as shown by \citet{Winter:2012yq}. $L_{\rm SF}$ measures the SFR using Equation~\ref{eq:sfr_ir}.

\begin{figure*}
\includegraphics{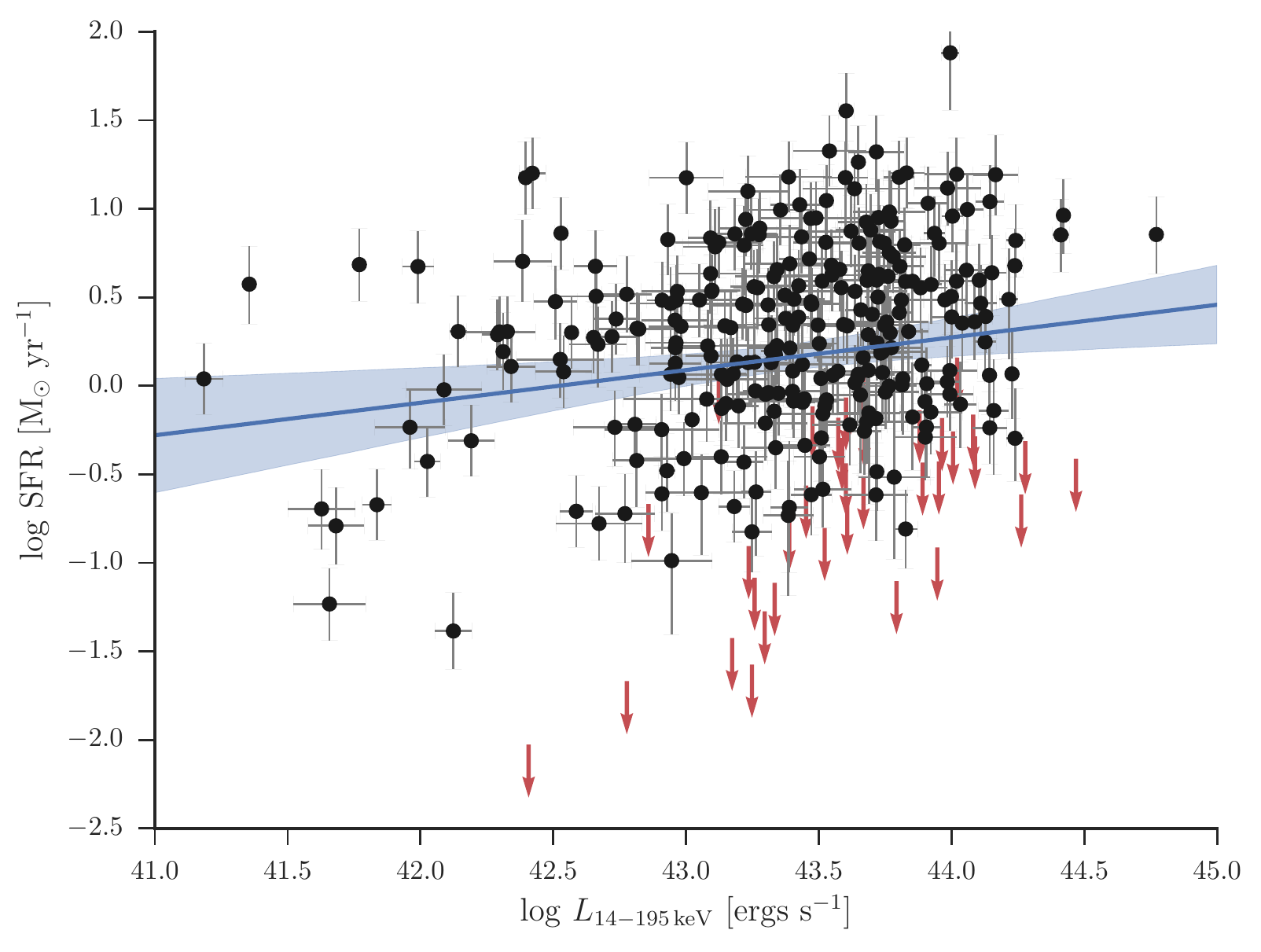}
\caption{Correlation between AGN luminosity as probed by $L_{\rm 14-195\,keV}$ and SFR. Error bars on $L_{\rm14-195\,keV}$ correspond to the 90 percentile confidence interval from the 58 month BAT catalogue. Error bars on the SFR correspond to the 68 percentile confidence interval from our SED fitting. Downward pointing arrows plot 95 percent confidence upper limits on the SFR. The solid blue line with shading plots the best fit line and 95 percentile confidence interval for our AGN-SFR relationship from \textsc{linmix\_err}. No strong relationship exists between the AGN strength and global star formation. A color version of this figure is available in the online publication.\label{fig:sfr_lbat_correlation}}
\end{figure*}

Figure~\ref{fig:sfr_lbat_correlation} plots the correlation between the SFR and $L_{\rm 14-195 keV}$ for the \herschel-BAT AGN. Black points with error bars indicate all of our detected sources, while red arrows show 95 per cent confidence upper limits for objects with less four SED points. We determine parameters for the following linear model:

\begin{equation}
\log {\rm SFR} = m\log L_{\rm 14-195\,keV} + b + \epsilon_{\rm int}\,,
\end{equation}

\noindent using \textsc{linmix\_err} as before. The blue line and shaded region in Figure~\ref{fig:sfr_lbat_correlation} represents the best-fit line and 95 percent confidence interval. We find a best fitting $m=0.18\pm0.07$, $b=-7.85\pm2.93$, and $\sigma^2=0.37\pm.04$ with a correlation coefficient of only 0.17$\pm$.06. Due to our large sample size, though, the probability for a null correlation is small with only a 0.25 per cent chance of measuring this correlation randomly.

\begin{figure*}
\includegraphics{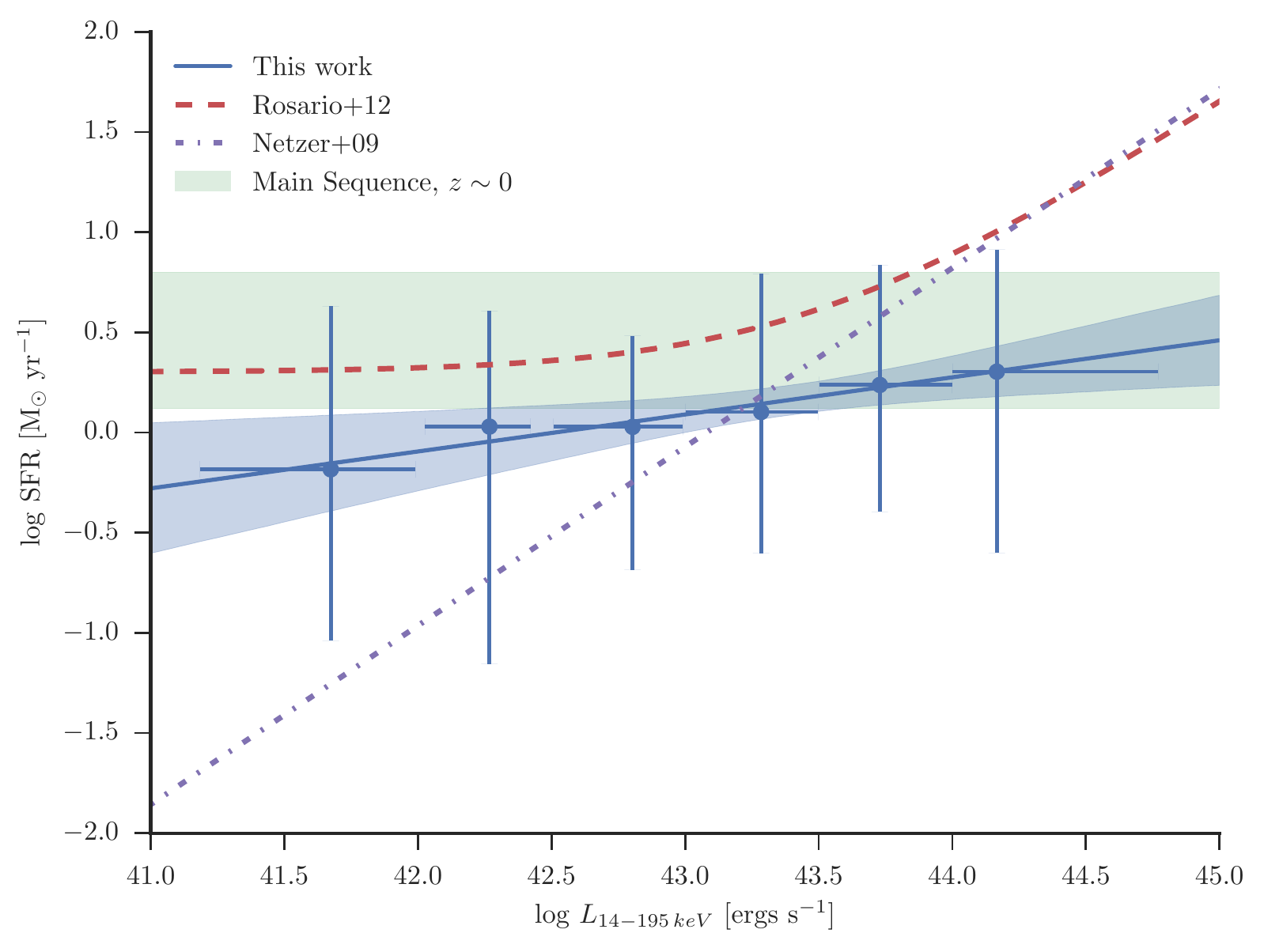}
\caption{Comparison between the measured SFR-$L_{\rm 14-195\,keV}$ from the \herschel-BAT sample (blue line with shaded region) and previous relationships from \citet{Netzer:2009lr} (dotted line) and \citet{Rosario:2012fr} (red dashed line). The points with error bars are the median SFRs binned by $L_{\rm 14-195\,keV}$ from the \herschel-BAT sample. Error bars represent the 68 per cent confidence interval.  The green shaded region represents the expected SFR for a main sequence galaxy in the local universe at a stellar mass of $10^{10.5}$ \msun. A color version of this figure is available in the online publication.\label{fig:sfr_lbat_correlation_lit}}
\end{figure*}

The scatter in the relationship is high, covering nearly two orders of magnitude in SFR for a given AGN luminosity. In Figure~\ref{fig:sfr_lbat_correlation_lit}, we compare our correlation with previous measured relationships at low redshift from \citet{Rosario:2012fr} (red dashed line) and \citet{Netzer:2009lr} (black dotted line). The blue points show the median $\log$ SFR in bins of 14--195 keV luminosity with error bars indicating the 16th and 84th percentiles. We further display as the green shaded region the expected SFR for main sequence galaxies with a stellar mass of $10^{10.5}$ \msun, equal to the average stellar mass of the \herschel-BAT sample. The shaded region spans the range of SFRs using different estimates of the main sequence from the literature including \citet{Shimizu:2015xo}, \citet{Schreiber:2015eu}, and \citet{Speagle:2014rc}. This reiterates our result from \citet{Shimizu:2015xo} that X-ray selected AGN largely reside in main sequence and below main sequence galaxies. 

Our correlation does not match with either the \citet{Rosario:2012fr} $z=0$ relationship (which is the same relationship presented in \citet{Lutz:2010kx} and \citet{Shao:2010fp}) or the \citet{Netzer:2009lr} relationship. Compared to \citet{Rosario:2012fr}, our correlation has a lower normalization and no evidence for an upturn at high luminosities. Compared to \citet{Netzer:2009lr}, our correlation has a flatter slope. The slope of the \citet{Netzer:2009lr} relationship is 0.8, which also coincides with the slope of the \citet{Rosario:2012fr} relationship at high luminosities. 

As \citet{Rosario:2012fr} point out, the discrepancy between \citet{Netzer:2009lr} and studies which find a flatter slope, including this one, is the focus in \citet{Netzer:2009lr} on AGN-dominated galaxies. This selection criteria could have produced two effects: 1) selecting against highly star-forming, low luminosity AGN host galaxies and 2) an unaccounted for contribution from the AGN to the SFR estimate. \citet{Netzer:2009lr} specifically uses the 60 \micron{} luminosity as their proxy for the SFR, which we have shown in this work can be largely affected by AGN heated dust. 

The differences between our relationship and the $z=0$ relationship from \citet{Rosario:2012fr} is interesting given both are based on the sample parent sample of AGN from the \swift/BAT catalogue. While we use individual \herschel{} detections and SED fitting to determine the SFR, \citet{Rosario:2012fr} relied on the 60 \micron{} luminosity derived from the \textit{IRAS} survey (Faint Source Catalogue or SCANPI). Our sample as well imposed a redshift cutoff of $z<0.05$, while \citet{Rosario:2012fr} used all AGN with $z < 0.3$. 

Due to the use of a single 60 \micron{} luminosity to estimate the SFR, our lower normalization of the relationship could be related to the subtraction of the AGN component in our SED fitting. This could also explain the upturn seen in \citet{Rosario:2012fr}. At higher AGN luminosities, the AGN component will dominate over the host galaxy and contribute more to the monochromatic 60 \micron{} luminosity.

However, it is still possible that our sample did not extend to high enough AGN luminosity to reveal the upturn seen in \citet{Rosario:2012fr}. The upper error bars on our binned points do extend to the \citet{Rosario:2012fr} relationship making the difference between our relationship and \citet{Rosario:2012fr} only $\sim1\sigma$ although the 95 per cent confidence interval on our relationship does not overlap with \citet{Rosario:2012fr}. Nevertheless, perhaps with more sources at higher AGN luminosity, an upturn could be revealed. The highest AGN luminosities are where we expect the AGN to have the greatest influence and the strongest link to current star formation as seen in merger simulations \citep[e.g.][]{Di-Matteo:2005lr}. \citet{Treister:2012rt} further showed that major mergers trigger only the most luminous AGN while moderate luminosity AGN seem to be fueled by secular processes like disk instabilities. Therefore, perhaps the moderate luminosity AGN that comprise our sample are simply not probing the stage in an AGN host galaxy's lifetime where global star formation and nuclear activity are linked. 

At moderate luminosities, two different explanations have been proposed to explain the lack of a relationship seen between the strength of the AGN and SFR. Recently, \citet{Diamond-Stanic:2012rw} suggested that the AGN only influences star formation in the nuclear regions of galaxies and find that when restricting their measurements to only the inner 1kpc, a stronger relationship appears. \citet{Esquej:2014vl}, probing even smaller scales ($r < 100$ pc) using MIR interferometry, also found a nearly linear relationship between the nuclear SFR and SMBH accretion rate, and  \citet{LaMassa:2013hb}, utilizing the fixed aperture of the SDSS fiber, found that only within 1.7 kpc a positive relationship between the SFR and accretion rate occurs. These are all in agreement with hydrodynamical simulations \citep{Hopkins:2010kx, Thacker:2014pd} that predict an increasingly linear relationship as the star formation size scale decreases. Work is currently ongoing to spatially map the SFRs of the \herschel-BAT sample to determine whether the relationship strengthens near the nucleus.

The second explanation involves the varying timescales associated with star formation and accretion onto the SMBH. Measuring the SFR from the IR luminosity results in an average SFR over nearly 100 Myr \citep{Kennicutt:2012it}, while the X-ray luminosity is more aligned with the instantaneous AGN luminosity especially given the observation that the X-ray emission likely originates very near to the SMBH \citep[e.g.][]{Chen:2011lr}. Therefore, if the AGN luminosity can vary over 100 Myr, while the SFR is relatively stable, this will cause any intrinsic relationship between the two to weaken. \citet{Hickox:2014yq} explored this using a simple toy model for the Eddington ratio distribution for an AGN and a linear relationship between the AGN luminosity and SFR, finding that variability can largely explain the observed correlations between SFR and AGN luminosity. This model also successfully reproduced the upturns seen in \citet{Rosario:2012fr}.

Recently, \citet{Stanley:2015qy} measured the SFR-AGN luminosity relationship for over 2000 X-ray detected AGN across a range of redshifts and compared their results to the predictions of toy models. They found that at all redshifts, the relationship is quite flat which matches the model predictions that describe AGN host galaxies having a highly variable AGN luminosity and living in relatively normal star-forming galaxies. 

We add their data (from their Table 2) to our SFR-AGN luminosity plot to see if the local universe is broadly following the higher redshift universe that \citet{Stanley:2015qy} explored. Our results are directly comparable since they similarly used \herschel{} data and SED fitting with an AGN component to calculate SFRs in X-ray detected AGN. They further also used survival analysis to include galaxies with only upper limits in their averages and analysis.

\begin{figure}
\includegraphics[width=\columnwidth]{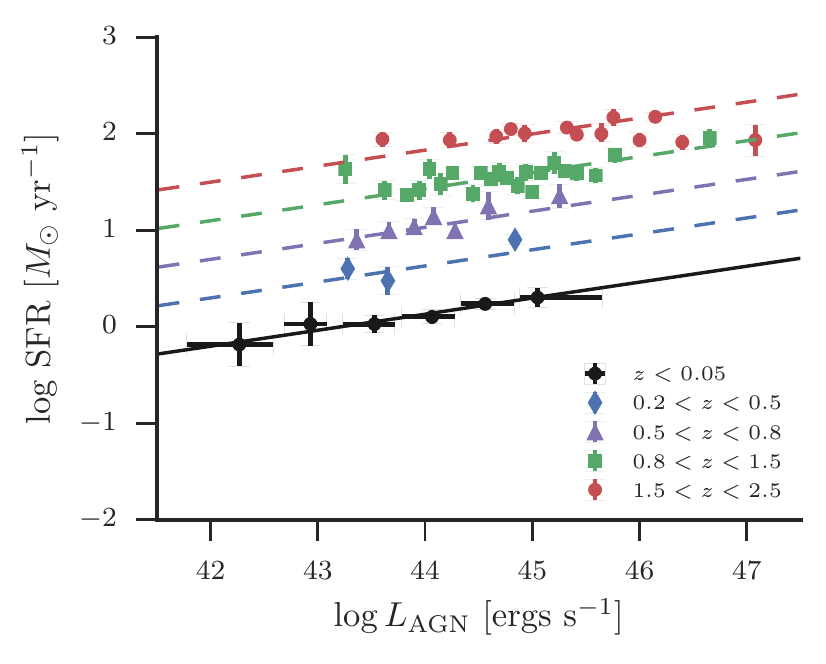}
\caption{\label{fig:sfr_lagn_correlation_stanley15} Evolution of the SFR-$L_{\rm AGN}$ relationship combining the results of this work (black points) and those of \citet{Stanley:2015qy} (colored points) for $z > 0.2$. The \herschel-BAT points were converted to $L_{\rm AGN}$ from $L_{\rm 14-195\,keV}$ using the calibration from \citet{Winter:2012yq}. Error bars on each point represent the standard error on the mean rather than the 68 per cent confidence interval as in Figure~\ref{fig:sfr_lbat_correlation_lit}. The black solid line shows our measured $z < 0.05$ SFR-$L_{\rm AGN}$ relationship. Each subsequent colored, dashed line is a simple scaling from $z < 0.05$ to higher redshifts by increasing the SFR by 0.5, 0.9, 1.3, and 1.7 dex.}
\end{figure}

Figure~\ref{fig:sfr_lagn_correlation_stanley15} shows the combined results from this work and \citet{Stanley:2015qy}. We converted our $L_{\rm 14-195\,keV}$ values to $L_{\rm AGN}$ using the calibration from \citet{Winter:2012yq} to properly compare the datasets. We also use the error on the mean rather than the 68 per cent confidence interval for the error bars on the SFR to follow \citet{Stanley:2015qy}. The addition of our $z<0.05$ points follow the same trends seen at higher redshifts. The average SFR of X-ray detected AGN host galaxies decreases from high to low redshift and within each redshift range the global SFR is only weakly related to $L_{\rm AGN}$. We show this in Figure~\ref{fig:sfr_lagn_correlation_stanley15} where the black solid line represents our measured relationship and each dashed line is our relationship but with 0.5, 0.9, 1.3, and 1.7 dex higher SFRs for the $0.2<z<0.5$, $0.5<z<0.8$, $0.8<z<1.5$, and $1.5<z<2.5$ redshift bins respectively. No formal fitting of the data was done only a shift by eye to test whether there is broad agreement up to $z\sim2.5$. Indeed, it seems that at least up to $z\sim1.5$, the same relationship as that found in the local universe can describe the connection between SFR and AGN luminosity. Only in the highest redshift bin and at the highest AGN luminosities does the local relationship appear to break down where SFRs are lower than expected.

The broad agreement between our results and those of \citet{Stanley:2015qy} indicates that the same properties and processes are leading to a weak or flat relationship between the global SFR and AGN strength. \citet{Stanley:2015qy} were able to fit their SFR-$L_{\rm AGN}$ relationships with an updated version of the \citet{Hickox:2014yq} variability model. They found that to reproduce the flatness, the faint end slope of the assumed Eddington ratio distribution needed to be steepened. This removed the strong upturns at high AGN luminosity which we also do not observe. Therefore, it is likely that the same model can accurately describe our local relationship.
 
If there is an intrinsically positive correlation between the SFR and AGN luminosity, whether on small spatial scales or long timescales, it can simply be explained by the availability of their common fuel, namely cold gas without invoking any AGN feedback. Large amounts of cold gas will spur both high accretion rates and high SFRs. Certainly, observations that a significant fraction of X-ray detected AGN lie below the star-forming main sequence \citep{Shimizu:2015xo, Matsuoka:2015fk, Mullaney:2015wn} are suggestive that the AGN is connected to quenching. AGN-driven outflows that are likely suppressing star formation have been observed in some samples \citep{Veilleux:2013qq, Cicone:2014ty, Genzel:2014rm, Tombesi:2015fj}; but in a recent study of 50 nearby \swift/BAT AGN, \citet{Stone:2016zl} found only four objects with evidence for a molecular outflow, strengthening the argument that AGN only contribute to driving gas out of the galaxy at high luminosity ($L_{\rm AGN} > 10^{45}$ ergs s$^{-1}$). Indeed, many of the sources where AGN driven outflows have been detected occur in powerful AGN \citep[e.g.][]{Cano-Diaz:2012en,Brusa:2015qk,Cresci:2015tk,Perna:2015rd}. 

Instead it is equally possible that at least some of our galaxies are going through passive quenching, whereby gas is simply being slowly depleted through star-formation and star-formation driven outflows after accretion onto the galaxy has slowed or been shut off \citep{Peng:2015yg}. This scenario agrees with the so-called ``bathtub'' model for galaxy evolution proposed in \citet{Lilly:2013pd}. The fact we observe a large fraction of AGN host galaxies in transition would be a consequence that AGN activity is more likely in galaxies with ongoing star formation, but there is a delay between the peak of star formation and accretion onto the SMBH \citep{Davies:2007kx, Sani:2012vn, Schawinski:2014cr} due to the longer timescales necessary for gas to reach the central SMBH. 

It is clear then that the situation involving the connection between an AGN and star formation in its host galaxy is quite complicated. Using our well-constrained IR luminosities and ultra-hard X-ray luminosities, we find a lack of a relationship between the SFR and AGN strength. Because our sample is not extending to quasar luminosities, though, we might be missing the phase of an AGN's lifetime where the strongest link between star formation and AGN luminosity is thought to exist. Further, the AGN influence might only work over small physical scales and AGN variability might be smearing any intrinsic relationship. Future work will involve remedying two of these problems by combining our sample with low redshift quasar samples like the Palomar-Green quasar sample and utilizing the relatively high spatial resolution of the \herschel{} images to look for a connection at smaller scales. 

\section{Conclusions}
Using our high quality \herschel{} photometry from \citet{Melendez:2014yu} and \citet{Shimizu:2016qy} combined with archival \textit{WISE} 12 and 22 \micron{} photometry, we have constructed and modeled the SEDs for over 300 AGN. Our sample is unique given its nearly unbiased selection based on ultra-hard X-ray detection, as well as its local nature that eliminates possible biases and source confusion. The following is a summary of our results and conclusions.

\begin{enumerate}
\item After correcting for host galaxy contribution, we find a nearly linear correlation between the estimated 8--1000 \micron{} luminosity due to the AGN and the 14--195 keV luminosity, signifying our decompositions are accurate.

\item We determined relationships between various MIR and FIR colors as a proxy for the AGN contribution to the total 8--1000 \micron{} luminosity ($f_{\rm AGN}$), finding that the 22/70 and 22/160 colors follow the best relationship with $f_{\rm AGN}$. 

\item We investigated the AGN contribution to the 70 \micron{} emission of a galaxy and showed that for nearly 30 per cent of our sample, more than half of the 70 \micron{} flux is likely due to AGN heating.

\item We calculated a median dust mass of $10^{7.36}$ \msun{}, dust temperature of $\sim 24$ K, and SFR of 1.7 \msun{} yr$^{-1}$ for AGN host galaxies. In a comparison with a mass-matched sample of non-AGN, we find all three properties are systematically higher by a factor 2 for the dust mass, 2 K in dust temperature, and a factor of 3 in SFR. This can be explained as a consequence of high mass non-AGN samples largely being composed of quiescent early type galaxies.

\item Confident in our measures of the SFR, we find a nearly flat (slope $= 0.18\pm0.07$) relationship between the SFR and AGN luminosity. Our flat relationship is in contrast to previous low redshift studies that find either a nearly linear relationship or an upturn at high luminosities but in agreement with recent studies at higher redshift. If nuclear activity is related to global star formation, it is only on long timescales; however, another possible explanation is that the relationship strengthens only as the physical scales probed decrease.
\end{enumerate}

We are working to determine whether spatial decomposition combined with our SED decomposition can illuminate the exact nature of the relationship between the AGN and nuclear star formation. This will also further test our simple decomposition methods and help expand our knowledge of the intrinsic AGN infrared SED into the FIR regime. Finally, we are also in the process of comparing the star forming properties for Type 1 and Type 2 objects to test if the obscuration of the AGN makes any significant difference.

\section*{Acknowledgements}
The authors thank the referee, Dr. James Mullaney, for his careful reading and thoughtful comments and suggestions that improved this manuscript. T.T.S. thanks Prof. Sylvain Veilleux and Dr. Dieter Lutz for helpful discussions and comments that improved this work. T.T.S. acknowledges support from a Graduate Dean's Dissertation Fellowship through the University of Maryland. We gratefully acknowledge support from NASA through Herschel contracts 1436909 (R.F.M., M.M., and T.T.S.), 1427288 (A.J.B.), and 1447980 (L.L.C.) This research has made use of the NASA/IPAC Extragalactic Database (NED), which is operated by the Jet Propulsion Laboratory, California Institute of Technology, under contract with the National Aeronautics and Space Administration. This publication makes use of data products from the Wide-field Infrared Survey Explorer, which is a joint project of the University of California, Los Angeles, and the Jet Propulsion Laboratory/California Institute of Technology, and NEOWISE, which is a project of the Jet Propulsion Laboratory/California Institute of Technology. WISE and NEOWISE are funded by the National Aeronautics and Space Administration. This research made use of Astropy, a community-developed core Python package for Astronomy \citep{Astropy:2013ek}. All figures in this publication were created with the Python package \textsc{MATPLOTLIB} \citep{Hunter:2007}. 


\bibliographystyle{mnras}
\bibliography{/Users/ttshimiz/Dropbox/Research/my_bib} 



\appendix

\section{SED Parameter Estimation}\label{sec:sed_bayes}
To find the best fitting parameters in our SED modeling, instead of standard least squares analysis, we used a Bayesian framework along with Monte Carlo Markov Chains to probe the posterior probability distribution functions for each parameter. The Bayesian framework allows for robust estimates of the uncertainty and for explicit statements about prior knowledge of the parameters. It also makes it relatively easy to include information contained in the undetected photometry of the SED.

\subsection{Likelihood Representation}
The likelihood defines the probability of observing a set of data given a specific model. In SED fitting, this translates to the combined probability of measuring all the photometric data points in the observed SED given a model for the SED (whether based on templates or analytic models). The total likelihood can then be expressed as the product of the probabilities of observing each single photometric point:

\begin{equation}\label{eq:likelihood}
\mathcal{L}(F|M) = \prod_{i}P(F_i|M)\,,
\end{equation}

\noindent where $F$ is the set of photometric fluxes, $F_i$ and $M$ is the model. For our analysis, we assume the probability of our observations follows a Gaussian distribution with mean equal to $M$ and standard deviations equal to the measurement errors, $\sigma_{i}$.

\begin{equation}\label{eq:detected_prob}
P(F_i|M) = \frac{1}{\sqrt{2\pi\sigma_{i}^{2}}}\mathrm{exp}\left(\frac{-(F_i - M)^2}{2\sigma_i^2}\right)\,,
\end{equation}

Equation~\ref{eq:detected_prob} only defines the probability for detected fluxes. To use the information contained in the undetected photometry, $U_i$, we define a different probability under the assumption that all of the upper limits are 5$\sigma$. 

\begin{align}\label{eq:undetected_prob}
P(U_i|M) &= \int_{-\infty}^{U_i}\frac{1}{\sqrt{2\pi\sigma_{i}^{2}}}\mathrm{exp}\left(\frac{-(x - M)^2}{2\sigma_i^2}\right)dx \notag\\
&= \frac{1}{2}\left(1 + \mathrm{erf}\left[\frac{U_i - M}{\sigma_i\sqrt{2}}\right]\right)\,,
\end{align}

\noindent where $\sigma_i = \frac{U_i}{5}$ and erf is the standard error function. For numerical accuracy and simplicity, it is customary to minimize the negative log-likelihood. Supposing we have $N$ total SED points with $D$ detections and $D-N$ non-detections then the total negative log-likelihood combining Equations~\ref{eq:likelihood},~\ref{eq:detected_prob},~and~\ref{eq:undetected_prob} is:

\begin{align}\label{eq:log_like}
-\log\mathcal{L} = \frac{1}{2}&\sum_{i=0}^{D}\left[\log(2\pi\sigma_i^2) - \left(\frac{F_i - M}{\sigma_i}\right)^2\right] + \notag \\
&\sum_{j=0}^{D-N}\log\left[1 + \mathrm{erf}\left(\frac{U_j - M}{\sigma_j\sqrt{2}}\right)\right]\,,
\end{align}

It is important to recognize here how $M$ is calculated, no matter whether it represents a template or analytic model. Each data point in an SED is the observer-frame flux density measured over a defined wavelength range. Therefore, to determine the model flux densities we first redshifted the full rest-frame model SED into the observer frame using the known redshifts of all of our sources. This observer-frame SED was then convolved with each instrument filter transmission curve to produce model flux densities that can be accurately compared to the observed ones.

\subsection{Bayesian MCMC Analysis}
Within the Bayesian framework, the important probability is the probability of the model given the data at hand, i.e. the most probable SED model given the observed fluxes. This probability can be determined using Bayes theorem:

\begin{equation}\label{eq:bayes}
P(M|F) = \frac{P(F|M)P(M)}{P(F)}\,,
\end{equation}

\noindent and is known as the posterior probability distribution. $P(F|M)$ is proportional to the likelihood (Equation~\ref{eq:likelihood}), $P(M)$ codifies our prior knowledge about the model, and $P(F)$ is the model evidence and can be disregarded as a simple normalization term. 

The reader may notice that assuming a flat prior, $P(M) \propto 1$, reduces Equation~\ref{eq:bayes} to $P(M|F) \propto \mathcal{L}$ verifying our use of maximum likelihood in determining the best template models.

For the C12 model, we used flat priors for the dust temperature and dust mass. We placed conservative limits on both the dust temperature and $\log M_{\mathrm dust}$ to be between 1 and 100 K and 1 and 10 \msun respectively. Within the powerlaw component for the C12 model, we also used flat priors for the powerlaw slope between -5 and 5 and the log of the normalization between -10 and 10. Based on previous work attempting to measure the intrinsic AGN SED and modeling the dusty torus, we expect the SED to turnover anywhere in the range between 20-70 \micron. Thus, we imposed a Gaussian prior centered at 45 \micron{} with a standard deviation of 20 \micron. We found that using this prior resulted in better and more realistic fits to the SEDs compared to a flat prior.

We used the \textsc{PYTHON} package \textsc{emcee} \citep{Foreman-Mackey:2013lr} to perform MCMC and sample the posterior probability distribution function (Equation~\ref{eq:bayes}). \textsc{emcee} runs an implementation of the Affine-Invariant MCMC sampler from Goodman \& Weare 2010. Instead of one single MCMC chain, it samples the posterior PDF with multiple ``walkers'', each with their own chain. For our analysis, we used 50 walkers that each produced a 1000 step chain. To allow for each chain to stabilize and move away from the initial guesses for the parameters, we imposed a 200 step ``burn-in''. In total, this resulted in 40000 steps to define the full posterior PDF. 

To determine the best fit parameters, we first marginalized the posterior PDF over all other parameters and then calculated the median. All quoted uncertainties represent the 68 per cent confidence interval determined from the 16th and 84th percentile of the marginalized posterior PDF.

For sources with less than four detections in their SED, we fixed the dust temperature to 25 K since the peak of the FIR bump is no longer constrained. In these cases, we only obtained upper limits on the dust mass by calculating the upper 95th percentile of the marginalized PDF.

\section{Comparison between different models}\label{sec:comp_models}
In this section we compare the results for the total luminosity, IR AGN luminosity, and star forming luminosity between the C12 model and two other models to decompose the SED. We exclude from this analysis objects which were only detected by \herschel{} in less than one waveband given the strong uncertainties associated with their properties. We also exclude as before the 6 radio-loud objects and Mrk 3 (for lack of WISE data). 

\subsection{DecompIR model}
Besides analytic models, another popular method is the use of template SEDs. Templates are constructed based on well-sampled SEDs of large samples of galaxies and usually parameterized according to a known property such as infrared luminosity.

For galaxies known to host an AGN, recent studies have turned to the \textsc{DecompIR} \citep{Mullaney:2011yq} templates. \textsc{DecompIR} consists of five host galaxy templates that span the IR color and luminosity range of the original \citet{Brandl:2006kx} starburst galaxy templates. \citet{Mullaney:2011yq} constructed the AGN templates based on a subsample of the \swift/BAT AGN which had AGN dominated \textit{Spitzer}/IRS spectra determined by the equivalent width of the 11.3 \micron{} feature being $<0.03$ \micron. The \textit{Spitzer}/IRS spectra were combined with \textit{IRAS} photometry at 60 and 100 \micron{} to define the ``intrinsic'' AGN SED from 6--100 \micron. 

\citet{Mullaney:2011yq} created three different AGN templates: one based only on high AGN luminosity objects, low AGN luminosity objects, and a median of the entire sample. For this work, we only consider the median AGN template given our SEDs only contain two points in the MIR where AGN-related emission is expected to dominate. 

\subsection{Dale et al 2014 model}
The third model we chose to test on our sample is the semi-empirical templates from \citet[][hereafter D14]{Dale:2014yq}. These templates also contain two components, one for dust emission in the host galaxy and one for the AGN. The host galaxy components were built from an updated version of the \citet{Dale:2002ty} model. Each component represents an SED produced using a different value of $\alpha_{\mathrm SF}$, which is the powerlaw slope of the intensity distribution for the interstellar radiation field that is heating the dust. These SEDs contain a mixture of emission from PAHs, small stochastically heated grains, and thermally radiating large grains.

For the AGN component, D14 chose the median SED of the Palomar-Green quasars from \citet{Shi:2013vn} citing the care with which any star-forming component was removed and the prominence of several AGN related MIR features such as the [OIV] fine structure line and the broad 10 and 18 \micron{} silicate emission bumps. At long wavelengths the AGN template falls as a blackbody.

Instead of two separate templates for the AGN and host galaxy, D14 provided a single set of templates based on different combinations of $\alpha_{\mathrm SF}$ and $f_{\mathrm AGN}$, the fractional contribution of the AGN to the 5--20 \micron{} emission. In total there are 1365 templates that range in $\alpha_{\mathrm SF}=0.0625-4.0$ in 0.0625 intervals and $f_{\mathrm AGN} = 0-1$ in 0.05 intervals.

\subsection{SED Fitting for the template models}
For each template in a model set, Equation~\ref{eq:log_like} was minimized to determine the best fit normalization. We then chose the normalized template with the lowest $-\log\mathcal{L}$ as the best fitting model for a source. For the DecompIR set, this meant first simultaneously optimizing over a normalization for the AGN component and each host galaxy component, then choosing the combined template that resulted in the minimum $-\log\mathcal{L}$. For D14, this meant calculating $-\log\mathcal{L}$ over the entire set of $\alpha$ and $f_{AGN}$ templates.

Uncertainties using the maximum likelihood method were determined by generating 1000 simulated SEDs for each source. These data points in the simulated SEDs were calculated by assuming each detected point followed a Gaussian distribution with mean equal to the observed flux density and a standard deviation equal to the measured error. Each of the simulated SEDs were re-fit using the same method. The standard deviation on the set of best fit parameters from the simulated SEDs then was used as the uncertainty. In this way both statistical and systematic errors can be taken into account in assessing the reliability of our best fit parameters. 

\subsection{$L_{\rm IR}$ Comparison}
\begin{figure*}
\includegraphics[width=0.75\textwidth]{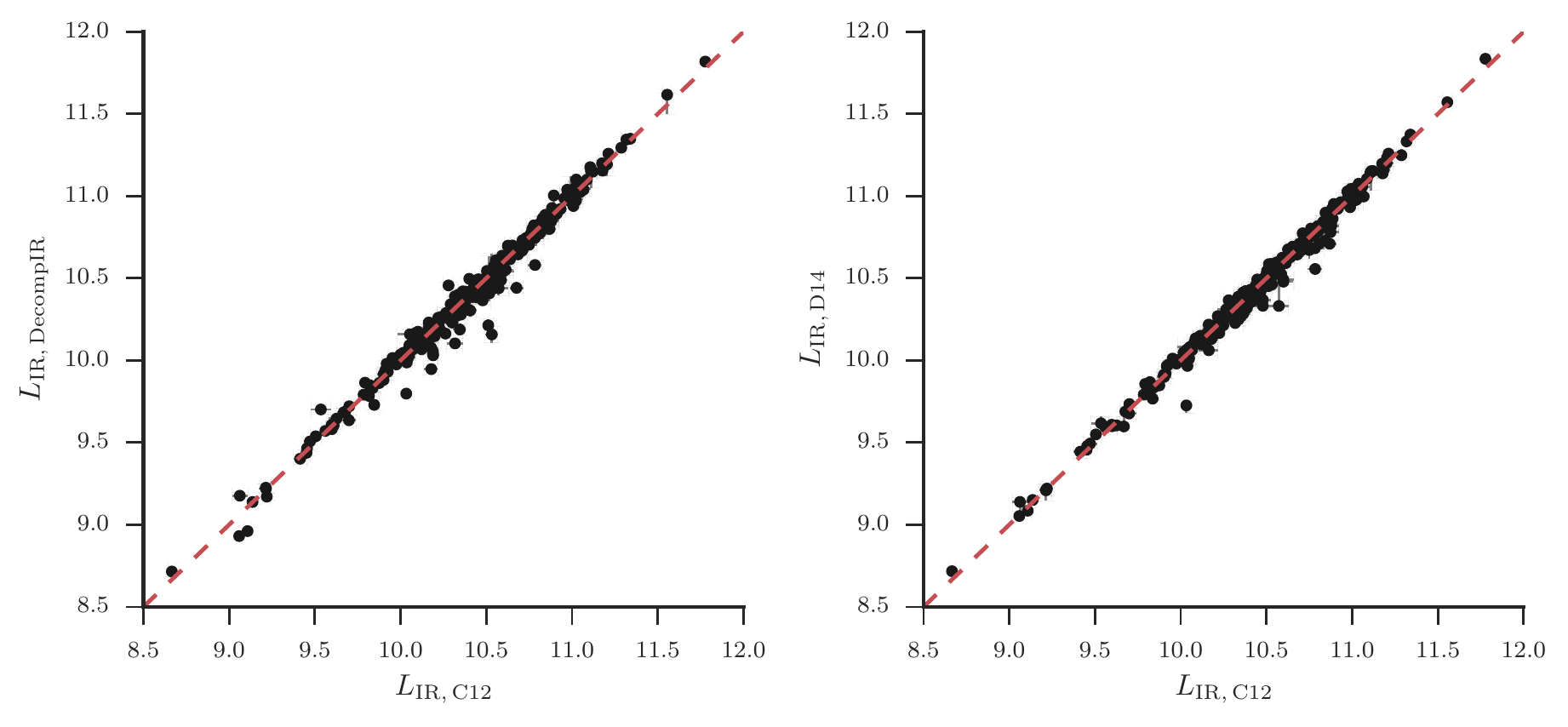}
\caption{Correlations of total $L_{\rm IR}$ between the C12 model and DecompIR and D14. The red dashed line indicates a 1-to-1 correspondence.\label{fig:lir_total_comp}}
\end{figure*}

The measured property that should be most model independent is the total infrared luminosity, $L_{\rm IR}$. Assuming each model was able to fit well the broadband SEDs and reproduce the observed photometry, $L_{\rm IR}$ is simply the total integrated energy underneath the SED, irregardless of how the SED is decomposed. In Figure~\ref{fig:lir_total_comp}, we plot the correlations between the C12 model and DecompIR and D14 models for $L_{\rm IR}$.  

The median $\log\,L_{\rm IR}$ for the C12, DecompIR, and D14 models are all 10.39 \lsun with a spread of 0.5 dex. Clearly, based on Figure~\ref{fig:lir_total_comp}, each model well fits the broadband SED of our sample with a small scatter around the 1-to-1 correspondence line. The average difference in the luminosities from each model is only $\sim0.01$ dex with a standard deviation between 0.05 and 0.08 dex, slightly higher than the median statistical uncertainty for the C12 model of 0.03 dex.  

\begin{figure*}
\includegraphics[width=0.75\textwidth]{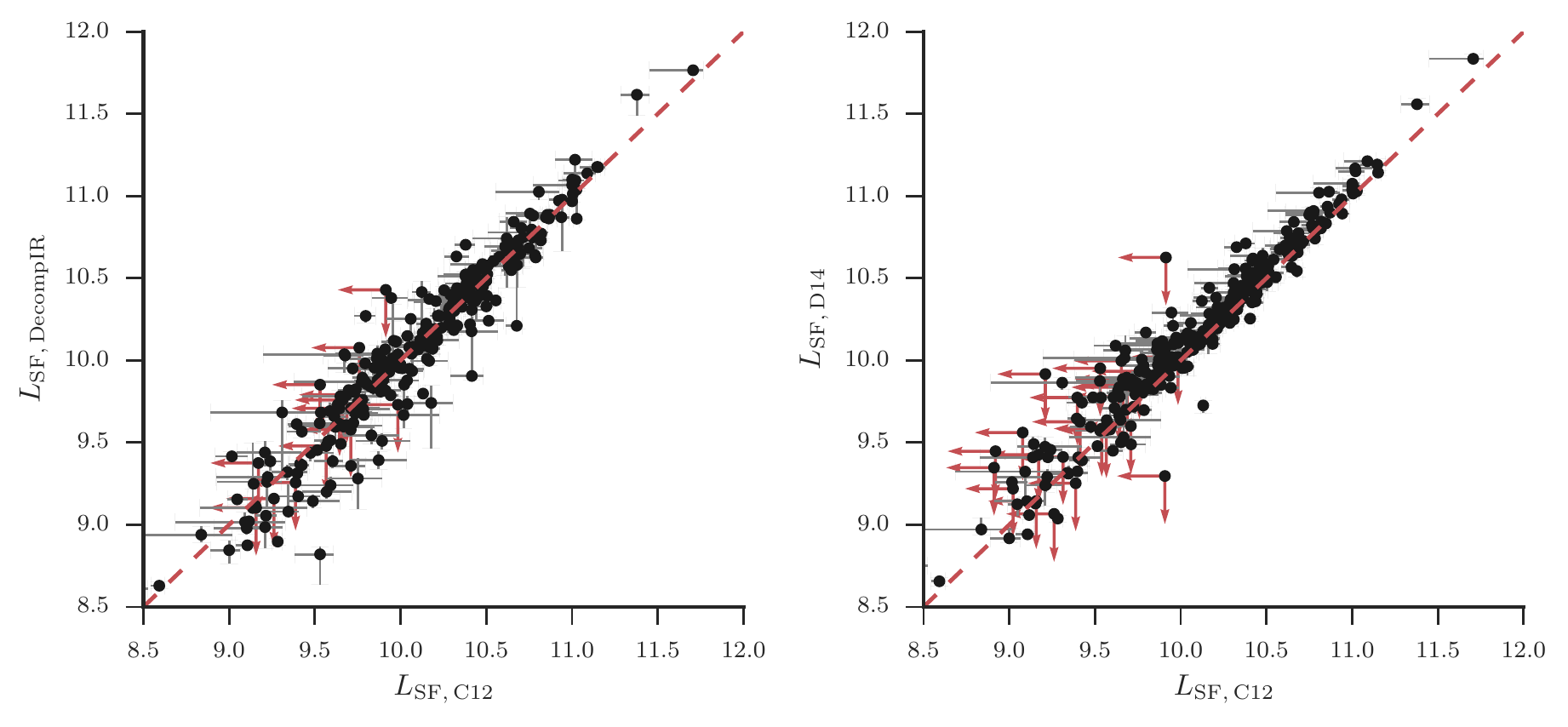}
\caption{Same as Figure~\ref{fig:lir_total_comp} but for $L_{\rm SF}$ \label{fig:lir_sf_comp}}
\end{figure*}

\begin{figure*}
\includegraphics[width=0.75\textwidth]{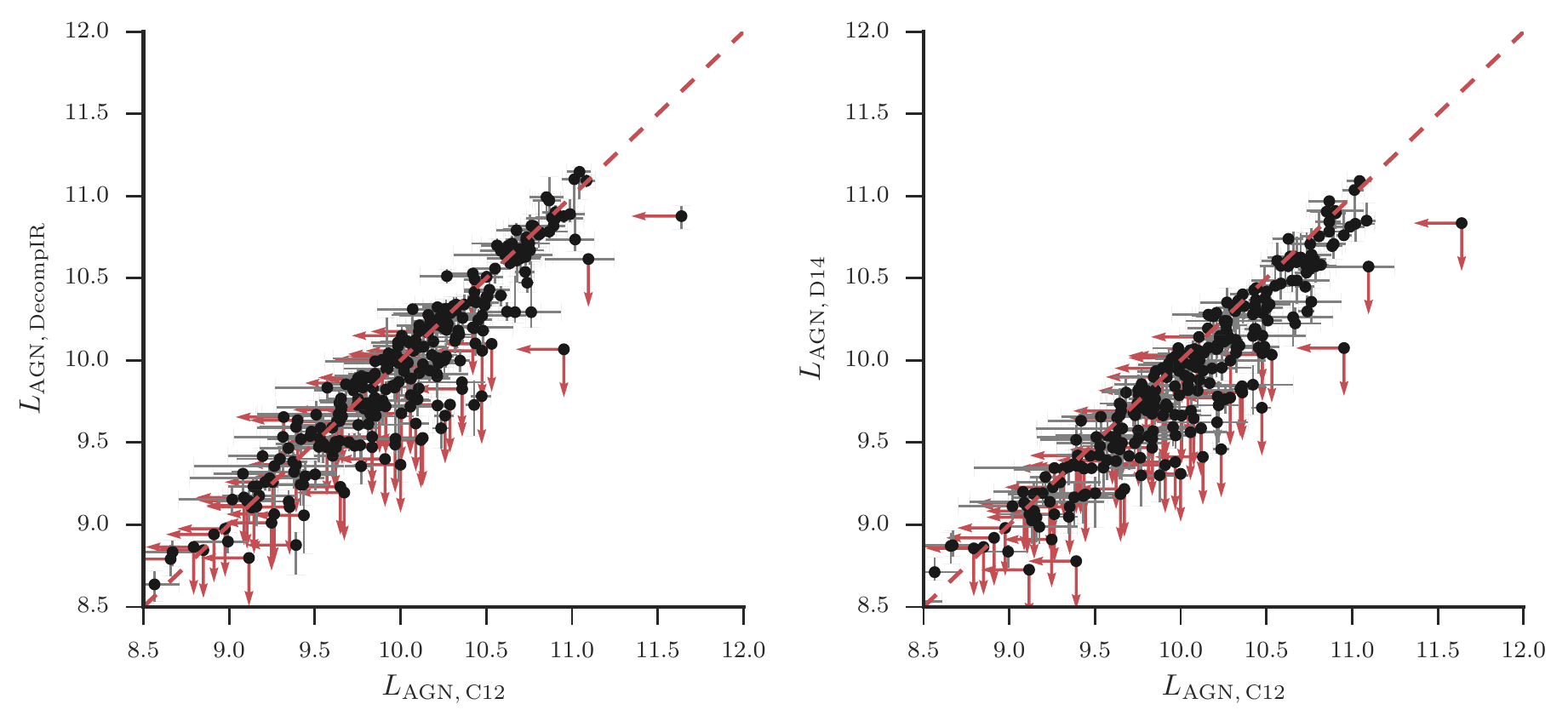}
\caption{Same as Figure~\ref{fig:lir_total_comp} but for $L_{\rm AGN,\,IR}$ \label{fig:lir_agn_comp}}
\end{figure*}

\subsection{$L_{\rm SF}$, $L_{\rm AGN, IR}$, and $f_{\rm AGN}$ Comparison}
Where the models begin to disagree more, is in the actual decomposition of the SED. In particular, we compare $L_{\rm SF}$, $L_{\rm AGN, IR}$, and$f_{\rm AGN}$. Calculating these parameters for the template models is relatively easy compared to the corrections we had to make for the C12 model. For the DecompIR template set, $L_{\rm IR}$, $L_{\rm SF}$, and $L_{\rm AGN, IR}$ were simply calculated from the best fit total, host galaxy, and AGN SEDs. For the D14 model set, $L_{\rm IR}$ was measured from the best fitting template and $L_{\rm SF}$ and $L_{\rm AGN, IR}$ were calculated based on the best fit $f_{\rm AGN}$.\footnote{\citet{Dale:2014yq} only provides $f_{\rm AGN}$ calculated between 5--15 \micron; however, D. Dale graciously provided the authors with $f_{\rm AGN}$ calculated between 8--1000 \micron{} through private communication in August 2015.} Uncertainties on all of these luminosities were determined with the same Monte Carlo method used to determine the uncertainties on the best fitting parameters.

\begin{figure*}
\includegraphics[width=0.8\textwidth]{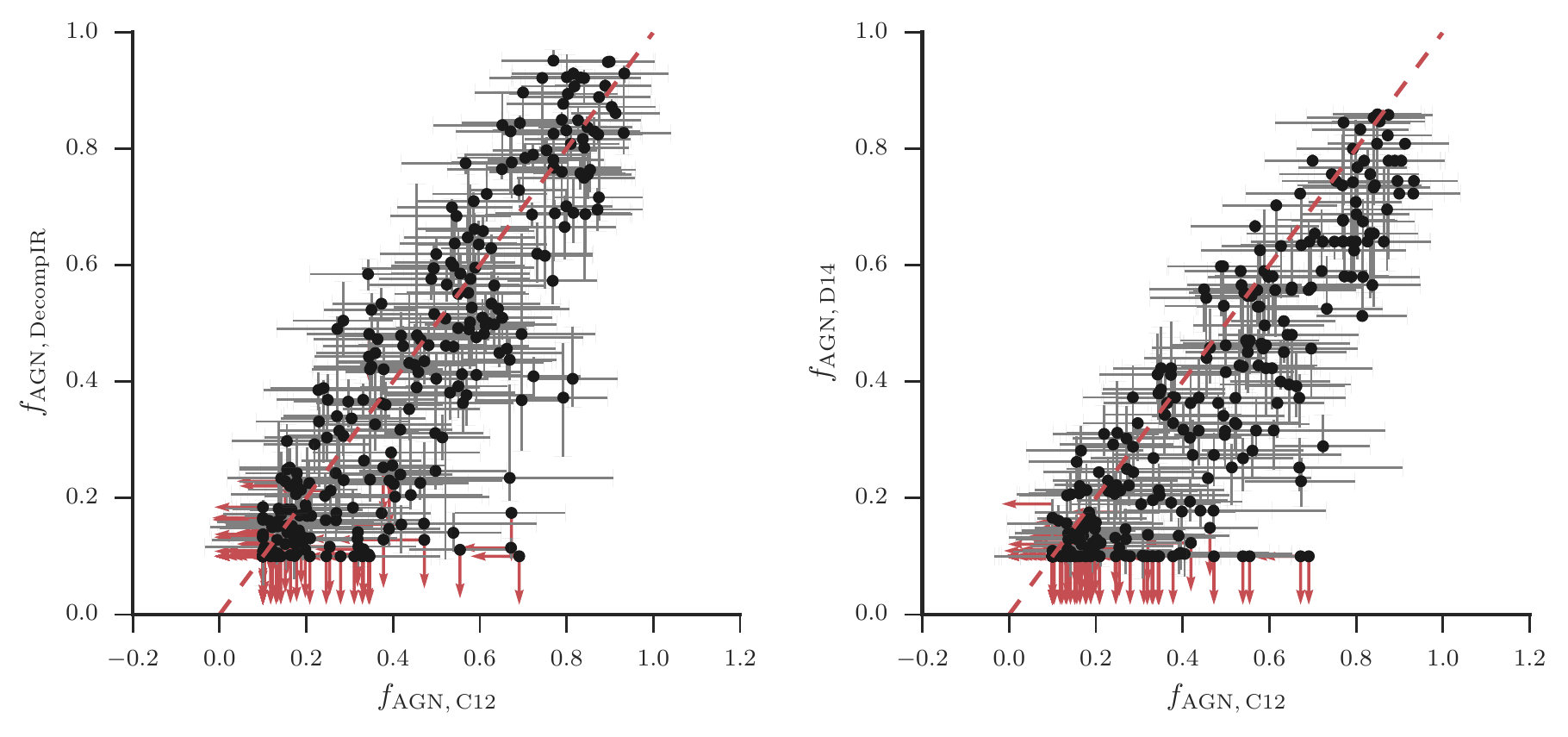}
\caption{Correlations of $f_{\rm AGN}$ from the three SED models.  Error bars represent the 68 per cent confidence interval. Upper limits are indicated as red downward pointing arrows. The red dashed line indicates a 1-to-1 correspondence. A color version of this figure is available in the online publication. \label{fig:agn_frac_comp}}
\end{figure*}

Figures~\ref{fig:lir_sf_comp},~\ref{fig:lir_agn_comp},~and~\ref{fig:agn_frac_comp} display the relationship between each model's $L_{\rm SF}$, $L_{\rm AGN, IR}$, and, $f_{\rm AGN}$. Out of the three parameters, $L_{\rm SF}$ is most consistent between the three models. The C12, DecompIR, and D14 models have a median $L_{\rm SF}$ of 10.12, 10.12, and 10.19 \lsun{} respectively. The average difference between C12 and DecompIR is only 0.006 dex with a standard deviation of 0.16 dex while the average difference between C12 and D14 is -0.07 dex with a standard deviation of 0.11 dex. The DecompIR and D14 are measuring slightly higher star-forming luminosities. Based on these differences, we assign a conservative uncertainty of 0.2 dex for $L_{\rm SF}$ to account for model dependence.

The comparison for $L_{\rm AGN, IR}$ shows increased scatter compared to the two previous luminosities. The median $L_{\rm AGN, IR}$ for the three models is 10.04, 9.9, and 9.85 \lsun{} for the C12, DecompIR, and D14 models respectively. The average difference between C12 and DecompIR is 0.10 dex with a standard deviation of 0.20 dex while the average difference between C12 and D14 is 0.14 dex with a standard deviation of 0.20 dex. Since the DecompIR and D14 were measuring higher $L_{\rm SF}$, it makes sense that they are measuring lower $L_{\rm AGN, IR}$ given the total infrared luminosities between the three models was identical. It is likely that the more flexible C12 model, which allows the PL component to extend to longer wavelengths, creates the discrepancy between it and the template based models. Our implementation of DecompIR and D14 adheres to a strict single template for the intrinsic AGN SED. Adding the offset and spread in the differences in quadrature indicates an uncertainty of 0.25 dex in the determination of $L_{\rm AGN, IR}$. 

The relatively large uncertainty in $L_{\rm AGN, IR}$ then leads to a large scatter in $f_{\rm AGN}$ as shown in Figure~\ref{fig:agn_frac_comp}. The average difference between C12 and DecompIR is only 0.02; however, the standard deviation is 0.13. The average difference between C12 and D14 is 0.1 with a standard deviation of 0.11. Given the spread in $f_{\rm AGN}$ is comparable to the average statistical uncertainty (0.14) from our C12 modeling, we do not add any more uncertainty onto our estimates. 

The 15 per cent uncertainty between the three models represents our general lack of knowledge about the details of decomposing broadband SEDs of AGN host galaxies. The estimated $f_{\rm AGN}$ is highly dependent on the assumed models of both the host galaxy and AGN, and currently at best we can only constrain to within 15 per cent. This influences then the calculations of both the SFR (based on $L_{\rm SF}$) and the IR portion of AGN luminosity ($L_{\rm AGN, IR}$). Studies relying on calculating the SFR of AGN host galaxies using the infrared need to take these discrepancies into account or else risk over-interpreting results based on broad SED decomposition. This also highlights the need for high angular resolution studies of AGN, to determine the true shape of the AGN IR SED.

\section{Comparison of $L_{\rm SF}$ with PAH measurements}\label{sec:pah_comp}
To test our estimates of the star-forming luminosity ($L_{\rm SF}$), we compare them with estimates of the star-forming luminosity calculated from the PAH 11.3 and 7.7 \micron{} luminosity. PAH features are thought to be caused by the vibrations of complex hydrocarbons \citep{Draine:1984fk, Puget:1989lr} and have been shown to be reliable tracers of the SFR \citep{Roussel:2001kx, Sargsyan:2009fj, Treyer:2010uq, Shipley:2016qy} even for AGN \citep{Diamond-Stanic:2010kl}. Thus, comparing our $L_{\rm SF}$ with the 11.3 and 7.7 \micron{} PAH luminosity provides a further check on our SED decomposition. 

To measure the 7.7 and 11.3 \micron{} luminosities we first searched the Cornell Atlas of Spitzer IRS Sources \citep[CASSIS;]{Lebouteiller:2011dk} for low-resolution Spitzer/IRS spectra of the \herschel-BAT AGN. We found 120/313 AGN on CASSIS and downloaded their reduced spectra using the default extraction method. Due to the different sizes of the Short-Low (SL) slit and Long-Low (LL) slit, the continuum of each can be offset mainly from extended emission not captured by the SL slit. Therefore, spectral orders were stitched together by fitting a line to the ends of each order and scaling to match the overlapping regions. SL2 and LL2 were first matched and scaled to SL1 and LL1 respectively. Then the combined SL1/SL2 spectra was matched to the combined LL1/LL2 spectra to produce the final full Spitzer/IRS low resolution spectra. 

We fit each spectra using small spectral windows centered on 7.7 and 11.3 \micron. Following \citet{Smith:2007lr} and the popular spectral fitting software package, PAHFIT, we modeled the PAH features using a Drude profile and any emission lines contained in the spectral windows as narrow Gaussians. The full-width-at-half-maximum (FWHM) and central wavelength of the emission lines were kept fixed at the values in Table 2 of \citet{Smith:2007lr}. The 7.7 and 11.3 \micron{} features are actually complexes with multiple subfeatures. PAHFIT models these with multiple Drude profiles chosen to reproduce the observed features and we follow this method using the values for the fractional contribution and FWHM as given in Table 3 of \citet{Smith:2007lr}. We model the continuum as a simple third-order polynomial. This allows the continuum to take on many shapes that is more necessary for AGN spectra, which can show broad silicate absorption and emission and varying continuum slopes \citep{Baum:2010kh, Spoon:2007oq, Wu:2009pt}. While silicate absorption is included within PAHFIT, silicate emission is absent prompting the need for our own modeling. Finally, to determine the best-fit model, we again use the Bayesian methodology and MCMC described in Appendix~\ref{sec:sed_bayes}. This allows us to treat the continuum and other nearby features as nuisance parameters and robustly determine uncertainties on the 7.7 and 11.3 \micron{} luminosities.

\begin{figure*}
\includegraphics{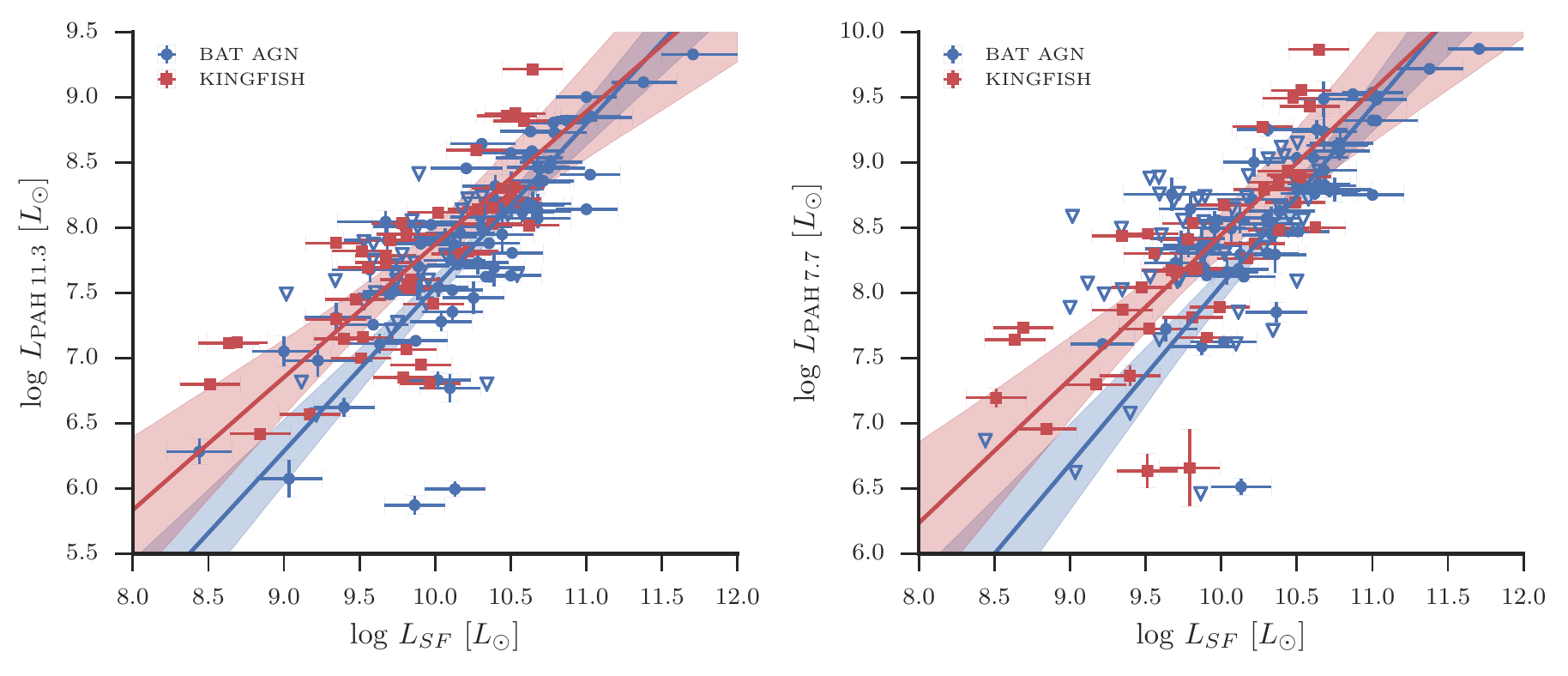}
\caption{Correlation between $L_{\rm SF}$ and the 11.3 (left) and 7.7 \micron{} (right) luminosity. Blue points correspond to the \herschel-BAT AGN and red squares correspond to the KINGFISH galaxies. Sources with upper limits on $L_{\rm PAH}$ are shown as open triangles. Measured correlations between $L_{\rm SF}$ and the 11.3 and 7.7 \micron{} luminosity are shown as solid blue and red lines for the \herschel-BAT AGN and KINGFISH samples respectively. The shaded region around the median correlations indicate the 95 per cent confidence interval. The broad consistency of the correlation for both the KINGFISH and \herschel-BAT samples shows that SFRs determined from $L_{\rm SF}$ will be comparable without any systematic bias. \label{fig:lsf_lpah}}
\end{figure*}

Figure~\ref{fig:lsf_lpah} shows the correlation between $L_{\rm SF}$ and $L_{\rm PAH}$ for both the 7.7 (right) and 11.3 \micron{} (left) complexes. We chose to use the KINGFISH galaxies (red squares) as a comparison here rather than HRS given the readily available PAH measurements. $L_{\rm SF}$ was determined using the C12 model in the same way as the HRS sample and $L_{\rm PAH}$ is from \citet{Smith:2007lr}. The $L_{\rm PAH}$ for KINGFISH were measured using PAHFIT, but given the large similarity with our method, we do not expect large differences.

Based on Figure~\ref{fig:lsf_lpah}, both the KINGFISH and \herschel-BAT samples seem to follow broadly the same correlation indicating we are neither over nor under estimating $L_{\rm SF}$ from our SED decomposition for the \herschel-BAT AGN. Using \textsc{LINMIX}, we measured the relationship for both samples and both PAH luminosities. The median correlations with 95 per cent confidence intervals are shown in Figure~\ref{fig:lsf_lpah} with the blue and red solid lines and shaded regions corresponding to the \herschel-BAT AGN and KINGFISH samples respectively. For the \herschel-BAT AGN we find slopes of $1.26\pm0.1$ (11.3 \micron) and $1.32\pm0.11$ (7.7 \micron) with correlation coefficients of 0.92 and 0.93 while for the KINGFISH galaxies we find slopes of $1.02\pm0.15$ and $1.11\pm0.16$ with correlation coefficients of 0.82 for both. In all four cases the probability of a null correlation is extremely small with the largest probability being $10^{-10}$. The \herschel-BAT AGN show a slightly steeper relationship than the KINGFISH galaxies, however they are still within $\sim1.5\sigma$ of each other. The steepness of the correlation for the \herschel-BAT AGN could be driven by the large number of upper limits for the PAH luminosities. Indeed, when excluding these from the analysis we find slopes of $1.13\pm0.09$ and $1.15\pm0.12$, much closer to the KINGFISH galaxies. 

For the purposes of this work, Figure~\ref{fig:lsf_lpah} indicates that our SED decomposition of the \herschel-BAT AGN has produced reliable estimates of the star-forming luminosity consistent with measurements of the PAH 11.3 and 7.7 \micron{} luminosity. A more detailed analysis on the exact correlation between broadband IR luminosity and PAH emission for AGN is beyond the scope of this Paper and has no bearing on our results.

\section{SED Figures}

\begin{figure*}
\centering
\includegraphics[width=\textwidth]{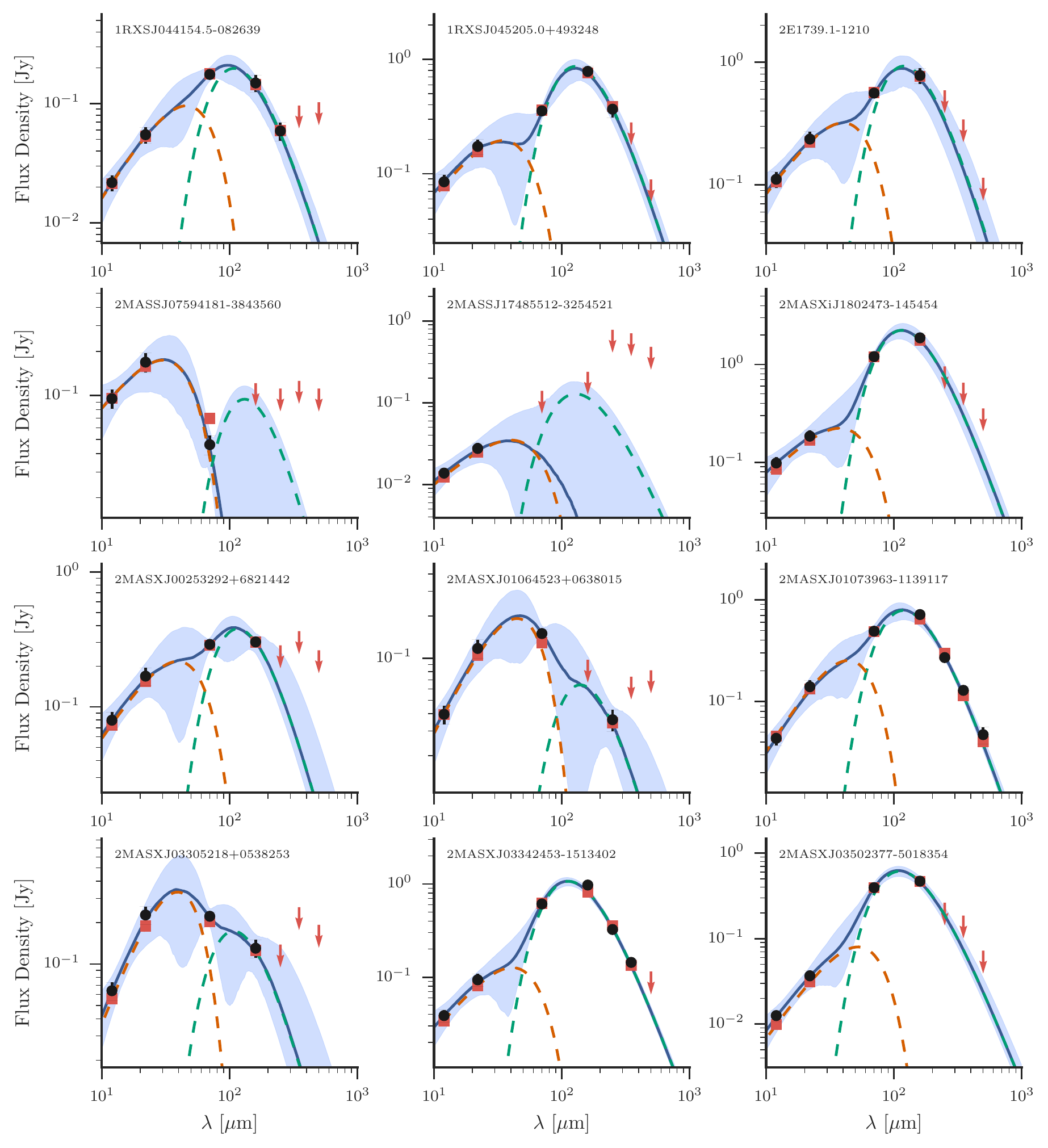}
\caption{Observed frame 12--500 \micron{} SEDs for all of the \herschel-BAT sample. Black points plot the observed flux densities with downward-pointing red arrows indicating 5$\sigma$ upper limits. The solid blue line and shaded region shows the best-fit C12 model with a 95 percent confidence interval. The red squares are the model flux densities after convolving the best-fit model SED with each instrument's transmission curve. The orange and green dashed lines show the best-fit PL and MBB components, respectively. \label{fig:seds}}
\end{figure*}

\bsp	
\label{lastpage}
\end{document}